\documentclass{article}

    \PassOptionsToPackage{numbers, compress}{natbib}

    \usepackage[preprint]{neurips2021}

\usepackage[utf8]{inputenc} 
\usepackage[T1]{fontenc}    
\usepackage{url}            
\usepackage{wrapfig}
\usepackage{booktabs}       
\newcommand{\ra}[1]{\renewcommand{\arraystretch}{#1}}
\usepackage{caption}
\usepackage{placeins}
\usepackage{amssymb}
\usepackage{pifont}
\usepackage{floatrow}
\usepackage{fancyvrb}
\usepackage{graphicx}
\usepackage{amsmath}
\usepackage{amsfonts}       
\usepackage{nicefrac}       
\usepackage{microtype}      
\usepackage{xcolor}         
\usepackage{xspace}         
\definecolor{magenta(dye)}{rgb}{0.79, 0.08, 0.48}

\definecolor{citecolor}{RGB}{34,139,34} 
\usepackage[pagebackref=true,breaklinks=true,colorlinks,
citecolor=citecolor,bookmarks=false]{hyperref}

\definecolor{bittersweet}{rgb}{1.0, 0.44, 0.37}

\definecolor{mygreen}{rgb}{0.29, 0.7, 0.48}

\definecolor{mygray}{gray}{0.4}
\newcommand{\cmark}{\color{mygray}\ding{51}}%
\newcommand{\xmark}{\color{mygray}\ding{55}}%

\title{\includegraphics[width=0.6cm]{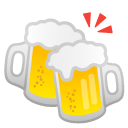} BEIR: A Heterogeneous Benchmark for Zero-shot Evaluation of Information Retrieval Models}

\author{Nandan Thakur, Nils Reimers, Andreas R\"uckl\'e\thanks{Contributions made prior to joining Amazon.}, Abhishek Srivastava, Iryna Gurevych \\
  Ubiquitous Knowledge Processing Lab (UKP-TUDA) \\
  Department of Computer Science, Technische Universit{\"a}t Darmstadt \\
  \href{https://www.informatik.tu-darmstadt.de/ukp/ukp_home/index.en.jsp}{\texttt{www.ukp.tu-darmstadt.de}}}

\newcommand{\custo}[1]{\textsc{\normalsize #1}}
\newcommand{\beir}{\custo{beir}\xspace}
\newfloatcommand{capbtabbox}{table}[][\FBwidth]

\begin{document}

\maketitle

\begin{abstract}
Existing neural information retrieval (IR) models have often been studied in homogeneous and narrow settings, which has considerably limited insights into their out-of-distribution (OOD) generalization capabilities. To address this, and to facilitate researchers to broadly evaluate the effectiveness of their models, we introduce \textbf{Be}nchmarking-\textbf{IR} (\textbf{\beir}), a robust and heterogeneous evaluation benchmark for information retrieval. We leverage a careful selection of 18 publicly available datasets from diverse text retrieval tasks and domains and evaluate 10 state-of-the-art retrieval systems including lexical, sparse, dense, late-interaction and re-ranking architectures on the \beir benchmark. 
Our results show BM25 is a robust baseline and re-ranking and late-interaction based models on average achieve the best zero-shot performances, however, at high computational costs. In contrast, dense and sparse-retrieval models are computationally more efficient but often underperform other approaches, highlighting the considerable room for improvement in their generalization capabilities. 
We hope this framework allows us to better evaluate and understand existing retrieval systems, and contributes to accelerating progress towards better robust and generalizable systems in the future. \beir is publicly available at \url{https://github.com/UKPLab/beir}.
\end{abstract}

\vspace{-3mm}
\section{Introduction}
\vspace{-1mm}

Major natural language processing (NLP) problems rely on a practical and efficient retrieval component as a first step to find relevant information. Challenging problems include open-domain question-answering \cite{chen-etal-2017-reading}, claim-verification \cite{thorne-etal-2018-fever}, duplicate question detection \cite{zhang2015multi}, and many more. Traditionally, retrieval has been dominated by lexical approaches like TF-IDF or BM25 \cite{bm25}. However, these approaches suffer from lexical gap \cite{berger2000bridging} and are able to only retrieve documents containing keywords present within the query. Further, lexical approaches treat queries and documents as bag-of-words by not taking word ordering into consideration.

Recently, deep learning and in particular pre-trained Transformer models like BERT \cite{devlin2018bert} have become popular in information retrieval \cite{lin2020pretrained}. These neural retrieval systems can be used in many fundamentally different ways to improve retrieval performance. We provide an brief overview of the systems in Section \ref{sec:neural_retrieval}. Many prior work train neural retrieval systems on large datasets like Natural Questions (NQ) \cite{47761} (133k training examples) or MS MARCO \cite{nguyen2016ms} (533k training examples), which both focus on passage retrieval given a question or short keyword-based query. In most prior work, approaches are afterward evaluated on the same dataset, where significant performance gains over lexical approaches like BM25 are demonstrated \cite{ding2020rocketqa, karpukhin-etal-2020-dense, nogueira2020passage}.

However, creating a large training corpus is often time-consuming and expensive and hence many retrieval systems are applied in a \textbf{zero-shot setup}, with no available training data to  train the system. So far, it is unclear how well existing trained neural models will perform for other text domains or textual retrieval tasks. Even more important, it is unclear how well different approaches, like sparse embeddings vs.\ dense embeddings, generalize to out-of-distribution data.

\begin{figure*}[t]
    \centering
    \begin{center}
        \vspace{-7mm}
        \includegraphics[trim=0 137 50 40,clip,width=\textwidth]{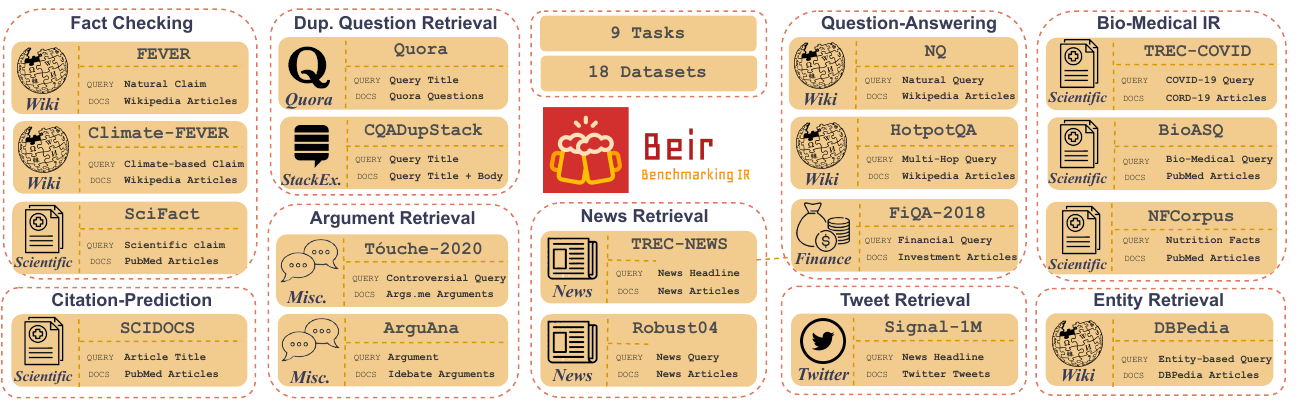}
        \caption{An overview of the diverse tasks and datasets in \beir benchmark. \vspace{-5mm}}
        \label{fig:beir-diagram}
    \end{center}
    \vspace*{-\baselineskip}
\end{figure*}

In this work, we present a novel robust and heterogeneous benchmark called \textbf{\beir} (\textbf{Be}nchmarking \textbf{IR}), comprising of 18 retrieval datasets for comparison and evaluation of model generalization. 
Prior retrieval benchmarks \citep{guo2020multireqa, petroni2020kilt} have issues of a comparatively narrow evaluation focusing either only on a single task, like question-answering, or on a certain domain. In \beir, we focus on \textbf{Diversity}, we include nine different retrieval tasks: Fact checking, citation prediction, duplicate question retrieval, argument retrieval, news retrieval, question answering, tweet retrieval, bio-medical IR, and entity retrieval. Further, we include datasets from diverse text domains, datasets that cover broad topics (like Wikipedia) and specialized topics (like COVID-19 publications), different text types (news articles vs.\ Tweets), datasets of various sizes (3.6k - 15M documents), and datasets with different query lengths (average query length between 3 and 192 words) and document lengths (average document length between 11 and 635 words).


We use \beir to evaluate \textbf{ten diverse retrieval methods} from five broad architectures: lexical, sparse, dense, late interaction, and re-ranking. From our analysis, we find that no single approach consistently outperforms other approaches on all datasets. Further, we notice that the in-domain performance of a model does not correlate well with its generalization capabilities: models fine-tuned with identical training data might generalize differently. In terms of efficiency, we find a trade-off between the performances and the computational cost: computationally expensive models, like re-ranking models and late interaction model perform the best. More efficient approaches e.g.\ based on dense or sparse embeddings can substantially underperform traditional lexical models like BM25. Overall, BM25 remains a strong baseline for zero-shot text retrieval.

Finally, we notice that there can be a strong lexical bias present in datasets included within the benchmark, likely as lexical models are pre-dominantly used during the annotation or creation of datasets. This can give an unfair disadvantage to non-lexical approaches. We analyze this for the TREC-COVID \cite{10.1145/3451964.3451965} dataset: We manually annotate the missing relevance judgements for the tested systems and see a significant performance improvement for non-lexical approaches. Hence, future work requires better unbiased datasets that allow a fair comparison for all types of retrieval systems.  

With \beir, we take an important step towards a single and unified benchmark to evaluate the zero-shot capabilities of retrieval systems. It allows to study when and why certain approaches perform well, and hopefully steers innovation to more robust retrieval systems. We release \beir and an integration of diverse retrieval systems and datasets in a well-documented, easy to use and extensible open-source package. \beir is model-agnostic, welcomes methods of all kinds, and also allows easy integration of new tasks and datasets. More details are available at \url{https://github.com/UKPLab/beir}.

\vspace{-2mm}
\section{Related Work and Background}
\vspace{-2mm}

To our knowledge, \beir is the first broad, zero-shot information retrieval benchmark. Existing works \cite{guo2020multireqa, petroni2020kilt} do not evaluate retrieval in a zero-shot setting in depth, they either focus over a single task, small corpora or on a certain domain. This setting hinders for investigation of model generalization across diverse set of domains and task types. MultiReQA \cite{guo2020multireqa} consists of eight Question-Answering (QA) datasets and evaluates sentence-level answer retrieval given a question. It only tests a single task and five out of eight datasets are from Wikipedia. Further, MultiReQA evaluates retrieval over rather small corpora: six out of eight tasks have less than 100k candidate sentences, which benefits dense retrieval over lexical as previously shown  \cite{reimers2020curse}. 
KILT \cite{petroni2020kilt} consists of five knowledge-intensive tasks including a total of eleven datasets. The tasks involve retrieval, but it is not the primary task. Further, KILT retrieves documents only from Wikipedia. 

\vspace{-3mm}
\subsection{Neural Retrieval}\label{sec:neural_retrieval}
\vspace{-2mm}

Information retrieval is the process of searching and returning relevant documents for a query from a collection. In our paper, we focus on text retrieval and use \textit{document} as a cover term for text of any length in the given collection and \textit{query} for the user input, which can be of any length as well. Traditionally, lexical approaches like TF-IDF and BM25 \cite{bm25} have dominated textual information retrieval. Recently, there is a strong interest in using neural networks to improve or replace these lexical approaches. In this section, we highlight a few neural-based approaches and we refer the reader to Lin et al. \cite{lin2020pretrained} for a recent survey in neural retrieval.

\textbf{Retriever-based} \quad Lexical approaches suffer from the lexical gap \cite{berger2000bridging}. To overcome this, earlier techniques proposed to improve lexical retrieval systems with neural networks. Sparse methods such as docT5query \cite{nogueira2019document} identified document expansion terms using a sequence-to-sequence model that generated possible queries for which the given document would be relevant. DeepCT \cite{10.1145/3397271.3401204} on the other hand used a BERT \cite{devlin-etal-2019-bert} model to learn relevant term weights in a document and generated a pseudo-document representation. Both  methods still rely on BM25 for the remaining parts. Similarly, SPARTA \cite{zhao-etal-2021-sparta} learned token-level contextualized representations with BERT and converted the document into an efficient inverse index.  More recently, dense retrieval approaches were proposed. They are capable of capturing semantic matches and try to overcome the (potential) lexical gap. Dense retrievers map queries and documents in a shared, dense vector space \cite{gillick2018endtoend}. This allowed the document representation to be pre-computed and indexed. A bi-encoder neural architecture based on pre-trained Transformers has shown strong performance for various open-domain question-answering tasks  \cite{guo2020multireqa, karpukhin-etal-2020-dense, liang2020embeddingbased, ma2021zeroshot}. This dense approach was recently extended by hybrid lexical-dense approaches which aims to combine the strengths of both approaches \cite{gao2020complementing, seo-etal-2019-real, luan2021sparse}. Another parallel line of work proposed an unsupervised domain-adaption approach \cite{liang2020embeddingbased,ma2021zeroshot} for training dense retrievers by generating synthetic queries on a target domain. Lastly, ColBERT \cite{10.1145/3397271.3401075} (Contextualized late interaction over BERT) computes multiple contextualized embeddings on a token level for queries and documents and uses an maximum-similarity function for retrieving relevant documents.

\textbf{Re-ranking-based} \quad Neural re-ranking approaches use the output of a first-stage retrieval system, often BM25, and re-ranks the documents to create a better comparison of the retrieved documents. Significant improvement in performance was achieved with the cross-attention mechanism  of BERT \cite{nogueira2020passage}. However, at a disadvantage of a high computational overhead \cite{reimers-2019-sentence-bert}.

\vspace{-3mm}
\section{The BEIR Benchmark} \label{sec_beir_benchmark}
\vspace{-2mm}

\beir aims to provide a one-stop zero-shot evaluation benchmark for all diverse retrieval tasks. To construct a comprehensive evaluation benchmark, the selection methodology is crucial to collect tasks and datasets with desired properties. For \beir, the methodology is motivated by the following three factors: (\emph{i}) \textbf{Diverse tasks}: Information retrieval is a versatile task and the lengths of queries and indexed documents can differ between tasks. Sometimes, queries are short, like a keyword, while in other cases, they can be long like a news article. Similarly, indexed documents can sometimes be long, and for other tasks, short like a tweet. (\emph{ii}) \textbf{Diverse domains}: Retrieval systems should be evaluated in various types of domains. From broad ones like News or Wikipedia, to highly specialized ones such as scientific publications in one particular field. Hence, we include domains which provide a representation of real-world problems and are diverse ranging from generic to specialized. (\emph{iii}) \textbf{Task difficulties}: Our benchmark is challenging and the \textit{difficulty} of a task included has to be sufficient. If a task is easily solved by any algorithm, it will not be useful to compare various models used for evaluation. We evaluated several tasks based on existing literature and selected popular tasks which we believe are recently developed, challenging and are not yet fully solved with existing approaches. (\emph{iv}) \textbf{Diverse annotation strategies}: Creating retrieval datasets are inherently complex and are subject to \textit{annotation biases} (see Section \ref{sec:biases-analysis} for details), which hinders a fair comparison of approaches. To reduce the impact of such biases, we selected datasets which have been created in many different ways: Some where annotated by crowd-workers, others by experts, and others are based on the feedback from large online communities.  

In total, we include 18 English zero-shot evaluation datasets from 9 heterogeneous retrieval tasks. As the majority of the evaluated approaches are trained on the MS MARCO \cite{nguyen2016ms} dataset, we also report performances on this dataset, but don't include the outcome in our zero-shot comparison. We would like to refer the reader to Appendix \ref{sec:datasets} where we motivate each one of the 9 retrieval tasks and 18 datasets in depth. Examples for each dataset are listed in Table \ref{tab:examples}. We additionally provide dataset licenses in Appendix \ref{sec:dataset_licenses}, and links to the datasets in Table \ref{tab:dataset_links}.

\begin{table*}[t!]
    \small
    \resizebox{\textwidth}{!}{\begin{tabular}{ l | l | l | c | c | c | c | c c c | c c }
        \toprule
         \multicolumn{1}{l}{\textbf{Split} ($\rightarrow$)} &
         \multicolumn{4}{c}{} &
         \multicolumn{1}{c}{\textbf{Train}}    &
         \multicolumn{1}{c}{\textbf{Dev}}    &
         \multicolumn{3}{c}{\textbf{Test}}   &
         \multicolumn{2}{c}{\textbf{Avg.~Word Lengths}} \\
         \cmidrule(lr){6-6}
         \cmidrule(lr){7-7}
         \cmidrule(lr){8-10}
         \cmidrule(lr){11-12}
           \textbf{Task ($\downarrow$)} &\textbf{Domain ($\downarrow$)} & \textbf{Dataset ($\downarrow$)} & \textbf{Title} & \textbf{Relevancy} & \textbf{\#Pairs} & \textbf{\#Query} & \textbf{\#Query} & \textbf{\#Corpus} & \textbf{Avg. D~/~Q } & \textbf{Query} & \textbf{Document} \\
         \midrule
    Passage-Retrieval    & Misc. & MS MARCO \cite{nguyen2016ms} & \xmark & Binary  & 532,761 &   ----  &   6,980   &   8,841,823      & 1.1 & 5.96  & 55.98  \\ \midrule[0.05pt] \midrule[0.05pt]
    Bio-Medical          & Bio-Medical & TREC-COVID \cite{10.1145/3451964.3451965}  & \cmark & 3-level &   ----    &   ----  & 50     & 171,332   & 493.5& 10.60 & 160.77 \\
    Information          & Bio-Medical & NFCorpus \cite{boteva2016}      & \cmark & 3-level & 110,575 &  324  & 323    & 3,633     & 38.2 & 3.30  & 232.26 \\
    Retrieval (IR)       & Bio-Medical & BioASQ \cite{tsatsaronis2015overview}       & \cmark & Binary  & 32,916 & ---- & 500    & 14,914,602& 4.7  & 8.05  & 202.61 \\ \midrule
    Question             & Wikipedia  & NQ  \cite{47761}           & \cmark & Binary  & 132,803  &   ----  & 3,452 & 2,681,468 & 1.2  & 9.16  & 78.88  \\
    Answering       & Wikipedia  & HotpotQA \cite{yang-etal-2018-hotpotqa}     & \cmark & Binary  & 170,000 & 5,447 & 7,405  & 5,233,329 & 2.0  & 17.61 & 46.30  \\
     (QA)           &Finance& FiQA-2018 \cite{10.1145/3184558.3192301}  & \xmark & Binary  & 14,166  &  500  & 648    & 57,638    & 2.6  & 10.77 & 132.32 \\ \midrule
    Tweet-Retrieval      &Twitter& Signal-1M (RT)  \cite{Signal1MRelatedTweetsRetrieval2018}    & \xmark & 3-level &   ----    &   ----  & 97     & 2,866,316 & 19.6 & 9.30  & 13.93  \\ \midrule
    News      &News& TREC-NEWS  \cite{soboroff2019trec}    & \cmark & 5-level &   ----    &   ----  & 57     & 594,977 & 19.6 & 11.14  & 634.79  \\
    Retrieval      &News& Robust04 \cite{96071}    & \xmark & 3-level &   ----    &   ----  & 249   & 528,155 & 69.9 & 15.27  & 466.40  \\ \midrule
    Argument       & Misc. & ArguAna  \cite{wachsmuth:2018a}    & \cmark & Binary  &   ----    &   ----  & 1,406  & 8,674     & 1.0  & 192.98& 166.80 \\
    Retrieval   & Misc. & Touch\'e-2020 \cite{stein:2020v} & \cmark & 3-level &   ----    &   ----  & 49     & 382,545   & 19.0 & 6.55  & 292.37 \\ \midrule
    Duplicate-Question   &StackEx.& CQADupStack \cite{hoogeveen2015cqadupstack}  & \cmark & Binary  &   ----    &   ----  & 13,145 & 457,199   & 1.4  & 8.59  & 129.09 \\
    Retrieval            & Quora &  Quora        & \xmark & Binary  &   ----    & 5,000 & 10,000 & 522,931   & 1.6  & 9.53  & 11.44  \\ \midrule
    Entity-Retrieval     & Wikipedia  &  DBPedia  \cite{Hasibi:2017:DVT}     & \cmark & 3-level &   ----    &   67  & 400    & 4,635,922 & 38.2 & 5.39  & 49.68  \\ \midrule
    Citation-Prediction  & Scientific&  SCIDOCS  \cite{cohan-etal-2020-specter}     & \cmark & Binary  &   ----    &   ----  & 1,000  & 25,657    & 4.9  & 9.38  & 176.19 \\ \midrule
                         & Wikipedia  &  FEVER \cite{thorne-etal-2018-fever}       & \cmark & Binary  & 140,085 & 6,666 & 6,666  & 5,416,568 & 1.2  & 8.13  & 84.76  \\ 
    Fact Checking        & Wikipedia  & Climate-FEVER \cite{diggelmann2020climatefever} & \cmark & Binary  &   ----    &   ----  & 1,535  & 5,416,593 & 3.0  & 20.13 & 84.76  \\
                         & Scientific & SciFact  \cite{wadden-etal-2020-fact}     & \cmark & Binary  &   920      &   ----  &  300   & 5,183     & 1.1  & 12.37 & 213.63  \\
    \bottomrule
    \end{tabular}}
    \vspace{-2mm} 
    \caption{\textbf{Statistics of datasets} in \beir benchmark. Few datasets contain documents without titles. Relevancy indicates the query-document relation: binary (relevant, non-relevant) or graded into sub-levels. Avg.~D/Q indicates the average relevant documents per query. \vspace{-5mm}}
    \label{tab:dataset_stats}
\end{table*}

Table \ref{tab:dataset_stats} summarizes the statistics of the datasets provided in \beir. A majority of datasets contain binary relevancy judgements, i.e.\ relevant or non-relevant, and a few contain fine-grained relevancy judgements. Some datasets contain few relevant documents for a query (< 2), while other datasets like TREC-COVID \cite{10.1145/3451964.3451965} can contain up to even 500 relevant documents for a query. Only 8 out of 19 datasets (including MS MARCO) have training data denoting the practical importance for zero-shot retrieval benchmarking. All datasets except ArguAna \cite{wachsmuth:2018a} have short queries (either a single sentence or 2-3 keywords). Figure \ref{fig:beir-diagram} shows an overview of the tasks and datasets in the \beir benchmark.

Information Retrieval (IR) is ubiquitous, there are lots of datasets available within each task and further even more tasks with retrieval. However, it is not feasible to include all datasets within the benchmark for evaluation. We tried to cover a balanced mixture of a wide range of tasks and datasets and paid importance not to overweight a specific task like question-answering. Future datasets can  easily be integrated in \beir, and existing models can be evaluated on any new dataset quickly. The \beir website will host an actively maintained leaderboard\footnote{\beir Leaderboard: \href{https://tinyurl.com/beir-leaderboard}{https://tinyurl.com/beir-leaderboard}} with all datasets and models.

\vspace{-3mm}
\subsection{Dataset and Diversity Analysis}\label{sec:dataset-diversity}
\vspace{-1mm}

\begin{figure*}[t]
\centering
\begin{center}
    \includegraphics[trim=5 10 180 100,clip,width=0.48\textwidth]{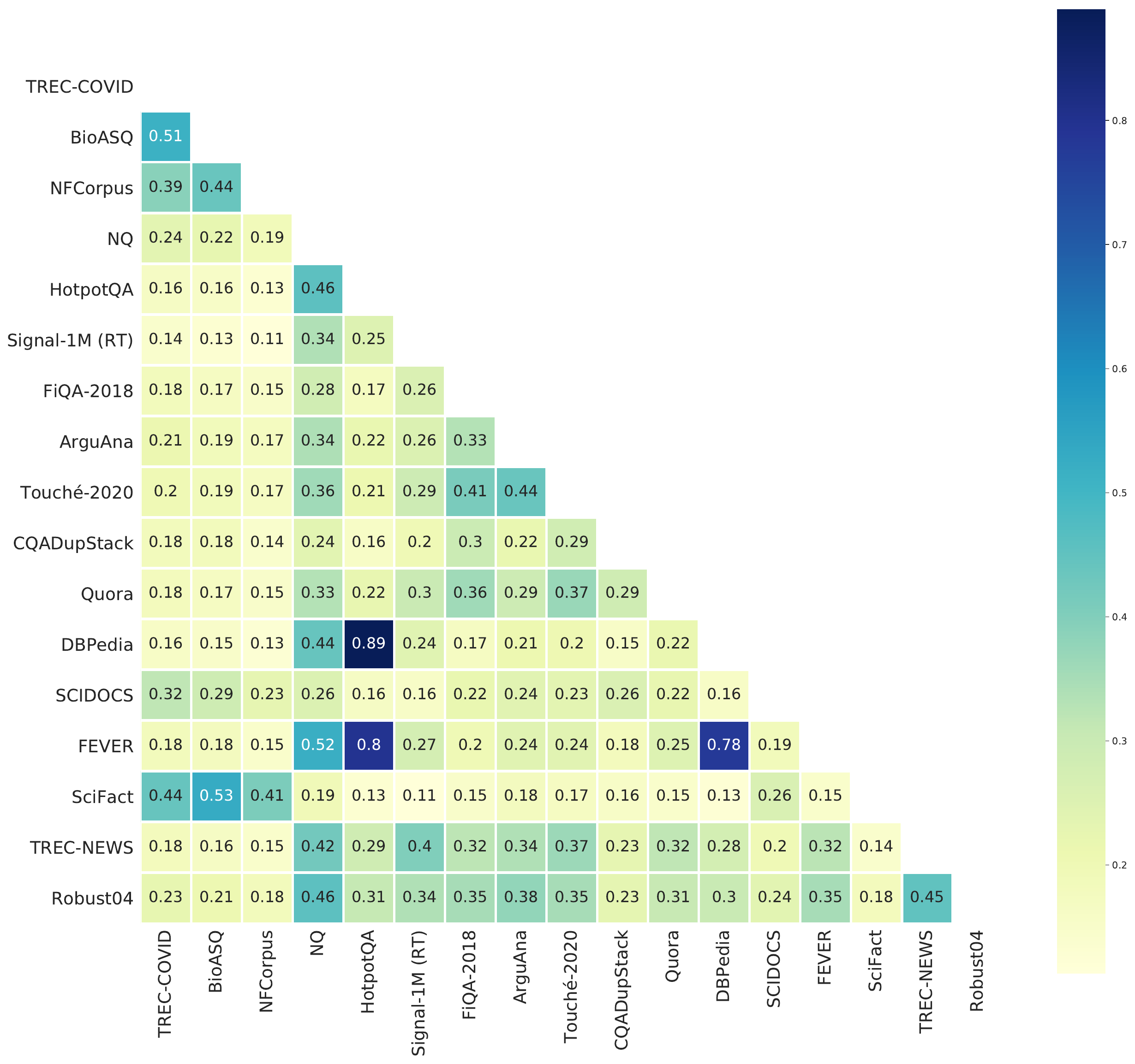}
    \includegraphics[trim=0 0 0 0,clip,width=0.5\textwidth]{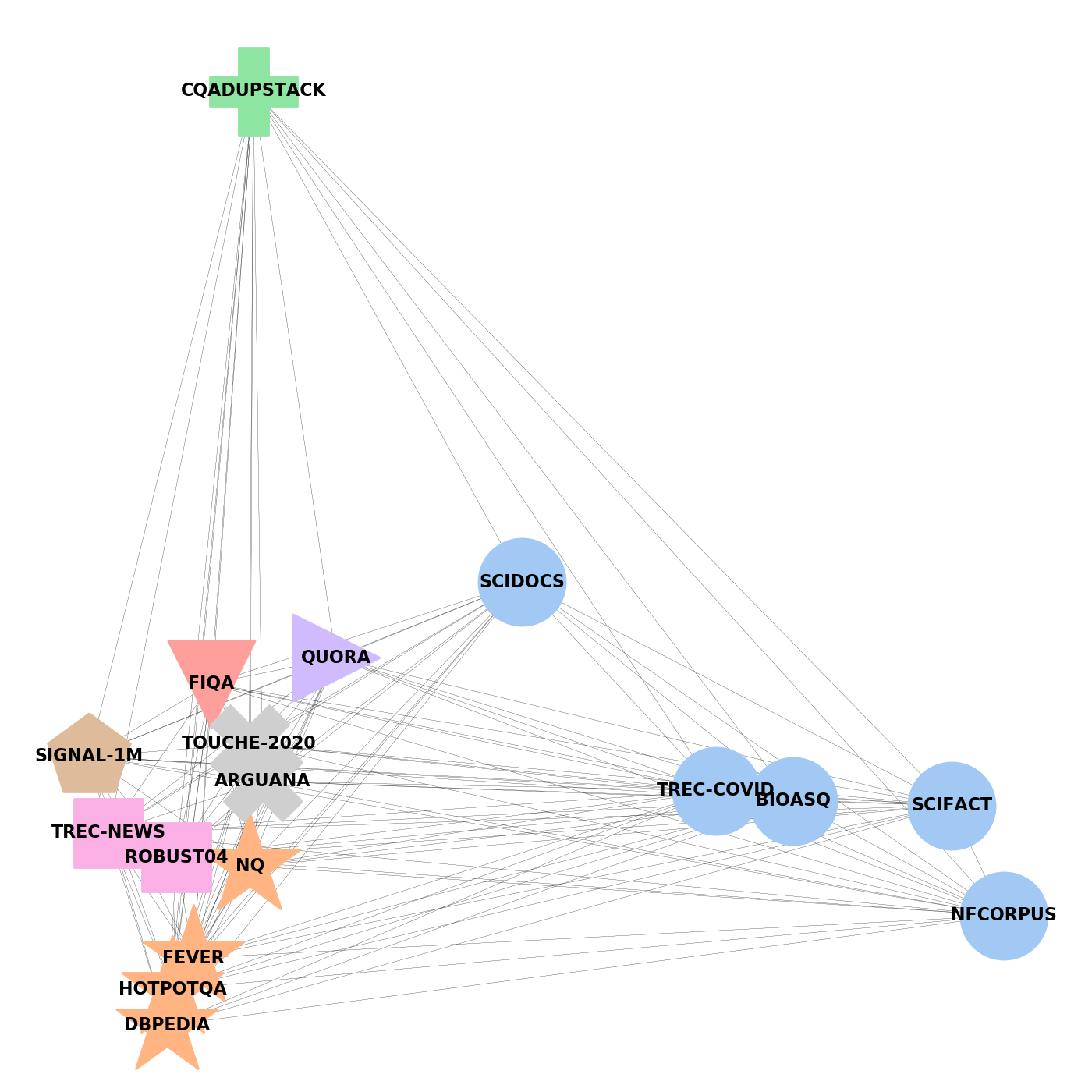}
    \caption{Domain overlap across each pairwise dataset in the \beir benchmark. Heatmap (left) shows the pairwise weighted jaccard similarity scores between \beir datasets. 2D representation (right) using a force-directed placement algorithm with NetworkX \cite{SciPyProceedings_11}. We color and mark datasets differently for different domains. \vspace{-5mm}}
    \label{fig:diversity-graph}
\end{center}
\vspace*{-\baselineskip}
\end{figure*}

The datasets present in \beir are selected from diverse domains ranging from Wikipedia, scientific publications, Twitter, news, to online user communities, and many more. To measure the diversity in domains, we compute the domain overlap between the pairwise datasets using a pairwise weighted Jaccard similarity \cite{ioffe2010improved} score on unigram word overlap between all dataset pairs. For more details on the theoretical formulation of the similarity score, please refer to Appendix \ref{sec:weighted_jaccard_similarity}. Figure \ref{fig:diversity-graph} shows a heatmap denoting the pairwise weighted jaccard scores and the clustered force-directed placement diagram. Nodes (or datasets) close in this graph have a high word overlap, while nodes far away in the graph have a low overlap. From Figure \ref{fig:diversity-graph}, we observe a rather low weighted Jaccard word overlap across different domains, indicating that \beir is a challenging benchmark where approaches must generalize well to diverse out-of-distribution domains.

\vspace{-2mm}
\subsection{BEIR Software and Framework}
\vspace{-1mm}

The \beir software\footnote{\beir Code \& documentation: \href{https://github.com/UKPLab/beir}{https://github.com/UKPLab/beir}} provides an is an easy to use Python framework (\texttt{pip install beir}) for model evaluation. It contains extensive wrappers to replicate experiments and evaluate models from well-known repositories including Sentence-Transformers \cite{reimers-2019-sentence-bert}, Transformers \cite{wolf-etal-2020-transformers}, Anserini \cite{10.1145/3077136.3080721}, DPR \cite{karpukhin-etal-2020-dense}, Elasticsearch, ColBERT \cite{10.1145/3397271.3401075}, and Universal Sentence Encoder \cite{yang2020multilingual}. This makes the software useful for both academia and industry. The software also provides you with all IR-based metrics from Precision, Recall, MAP (Mean Average Precision), MRR (Mean Reciprocal Rate) to nDCG (Normalised Cumulative Discount Gain) for any top-k hits. One can use the \beir benchmark for evaluating existing models on new retrieval datasets and for evaluating new models on the included datasets.

Datasets are often scattered online and are provided in various file-formats, making the evaluation of models on various datasets difficult. \beir introduces a standard format (corpus, queries and qrels) and converts existing datasets in this easy universal data format, allowing to evaluate faster on an increasing number of datasets. 

\vspace{-2mm}
\subsection{Evaluation Metric}\label{sec:evaluation_metric}
\vspace{-1mm}

Depending upon the nature and requirements of real-world applications, retrieval tasks can be either be precision or recall focused. To obtain comparable results across models and datasets in \beir, we argue that it is important to leverage a single evaluation metric that can be computed comparably across all tasks. Decision support metrics such as Precision and Recall which are both rank unaware are not suitable. Binary rank-aware metrics such as MRR (Mean Reciprocal Rate) and MAP (Mean Average Precision) fail to evaluate tasks with graded relevance judgements. We find that \textbf{Normalised Cumulative Discount Gain} (nDCG@k) provides a good balance suitable for both tasks involving binary and graded relevance judgements. We refer the reader to Wang et al. \cite{wang2013theoretical} for understanding the theoretical advantages of the metric. For our experiments, we utilize the Python interface of the official TREC evaluation tool \cite{VanGysel2018pytreceval} and compute nDCG@10 for all datasets.


\vspace{-2mm}
\section{Experimental Setup}
\vspace{-2mm}

We use \beir to compare diverse, recent, state-of-the-art retrieval architectures with a focus on transformer-based neural approaches. We evaluate on publicly available pre-trained checkpoints, which we provide in Table \ref{tab:model_links}. Due to the length limitations of transformer-based networks, we use only the first 512 word pieces within all documents in our experiments across all neural architectures.

We group the models based on their architecture: (\emph{i}) lexical, (\emph{ii}) sparse, (\emph{iii}) dense, (\emph{iv}) late-interaction, and (\emph{v}) re-ranking. Besides the included models, the \beir benchmark is model agnostic and in future different model configurations can be easily incorporated within the benchmark.

\textbf{(\emph{i}) Lexical Retrieval}: (\emph{a}) \textbf{BM25} \cite{bm25} is a commonly-used bag-of-words retrieval function based on token-matching between two high-dimensional sparse vectors with TF-IDF token weights. We use Anserini \cite{lin2016toward} with the default Lucene parameters (k=0.9 and b=0.4). We index the title (if available) and passage as separate fields for documents. In our leaderboard, we also tested Elasticsearch BM25 and Anserini + RM3 expansion, but found Anserini BM25 to perform the best. 

\textbf{(\emph{ii}) Sparse Retrieval}: (\emph{a}) \textbf{DeepCT} \cite{10.1145/3397271.3401204} uses a bert-base-uncased model trained on MS MARCO to learn the term weight frequencies (tf). It generates a pseudo-document with keywords multiplied with the learnt term-frequencies. We use the original setup of Dai and Callan \cite{10.1145/3397271.3401204} in combination with BM25 with default Anserini parameters which we empirically found to perform better over the tuned MS MARCO parameters. (\emph{b}) \textbf{SPARTA} \cite{zhao-etal-2021-sparta} computes similarity scores between the  non-contextualized query embeddings from BERT with the contextualized document embeddings. These scores can be pre-computed for a given document, which results in a 30k dimensional sparse vector. As the original implementation is not publicly available, we re-implemented the approach. We fine-tune a DistilBERT \cite{sanh2020distilbert} model on the MS MARCO dataset and use sparse-vectors with 2,000 non-zero entries. 
(\emph{c}) \textbf{DocT5query} \cite{nogueira2019doc2query} is a popular document expansion technique using a T5 (base) \cite{JMLR:v21:20-074} model trained on MS MARCO to generate synthetic queries and append them to the original document for lexical search. We replicate the setup of Nogueira and Lin \cite{nogueira2019doc2query} and generate 40 queries for each document and use BM25 with default Anserini parameters. 

\textbf{(\emph{iii}) Dense Retrieval}: 
(\emph{a}) \textbf{DPR} \cite{karpukhin-etal-2020-dense} is a two-tower bi-encoder trained with a single BM25 hard negative and in-batch negatives.  
We found the open-sourced Multi model to perform better over the single NQ model in our setting. The Multi-DPR model is a bert-base-uncased model trained on four QA datasets (including titles): NQ \cite{47761}, TriviaQA \cite{joshi-etal-2017-triviaqa}, WebQuestions \cite{berant-etal-2013-semantic} and CuratedTREC \cite{baudivs2015modeling}. 
(\emph{b}) \textbf{ANCE} \cite{xiong2020approximate} is a  bi-encoder constructing hard negatives from an Approximate Nearest Neighbor (ANN) index of the corpus, which in parallel updates to select hard negative training instances during fine-tuning of the model. We use the publicly available RoBERTa \cite{liu2019roberta} model trained on MS MARCO \cite{nguyen2016ms} for 600K steps for our experiments. 
(\emph{c}) \textbf{TAS-B} \cite{Hofstaetter2021_tasb_dense_retrieval} is a bi-encoder trained with Balanced Topic Aware Sampling using dual supervision from a cross-encoder and a ColBERT model. The model was trained with a combination of both a pairwise Margin-MSE \cite{hofstatter2021improving} loss and an in-batch negative loss function.
(\emph{d}) \textbf{GenQ}: is an unsupervised domain-adaption approach for dense retrieval models by training on synthetically generated data. First, we fine-tune a T5 (base) \cite{JMLR:v21:20-074} model on MS MARCO for 2 epochs. Then, for a target dataset we generate 5 queries for each document using a combination of top-k and nucleus-sampling (top-k: 25; top-p: 0.95). Due to resource constraints, we cap the maximum number of target documents in each dataset to 100K. For retrieval, we continue to fine-tune the TAS-B model using in-batch negatives on the synthetic queries and document pair data. Note, GenQ creates an independent model for each task.

\textbf{(\emph{iv}) Late-Interaction}: (\emph{a}) \textbf{ColBERT} \cite{10.1145/3397271.3401075} encodes and represents the query and passage into a bag of multiple contextualized token embeddings. The late-interactions are aggregated with sum of the max-pooling query term and a dot-product across all passage terms. We use the ColBERT model as a dense-retriever (end-to-end retrieval as defined \cite{10.1145/3397271.3401075}): first top-k candidates are retrieved using ANN with faiss \cite{JDH17} (faiss depth = 100) and ColBERT re-ranks by computing the late aggregated interactions. We train a bert-base-uncased model, with maximum sequence length of 300 on the MS MARCO dataset for 300K steps.

\textbf{(\emph{v}) Re-ranking model}: (\emph{a}) \textbf{BM25 + CE}\label{sec:electra_model} \cite{NEURIPS2020_3f5ee243} reranks the top-100 retrieved hits from a first-stage BM25 (Anserini) model. We evaluated 14 different cross-attentional re-ranking models that are publicly available on the HuggingFace model hub and found that a 6-layer, 384-h MiniLM \cite{NEURIPS2020_3f5ee243} cross-encoder model offers the best performance on MS MARCO. The model was trained on MS MARCO using a knowledge distillation setup with an ensemble of three teacher models: BERT-base, BERT-large, and ALBERT-large models following the setup in Hofstätter et al. \cite{hofstatter2021improving}.

\vspace{-3mm}
\section{Results and Analysis}
\vspace{-2mm}

\begin{figure}
\begin{floatrow}
\ffigbox{%
\includegraphics[trim=15 15 15 15,clip,width=0.48\textwidth]{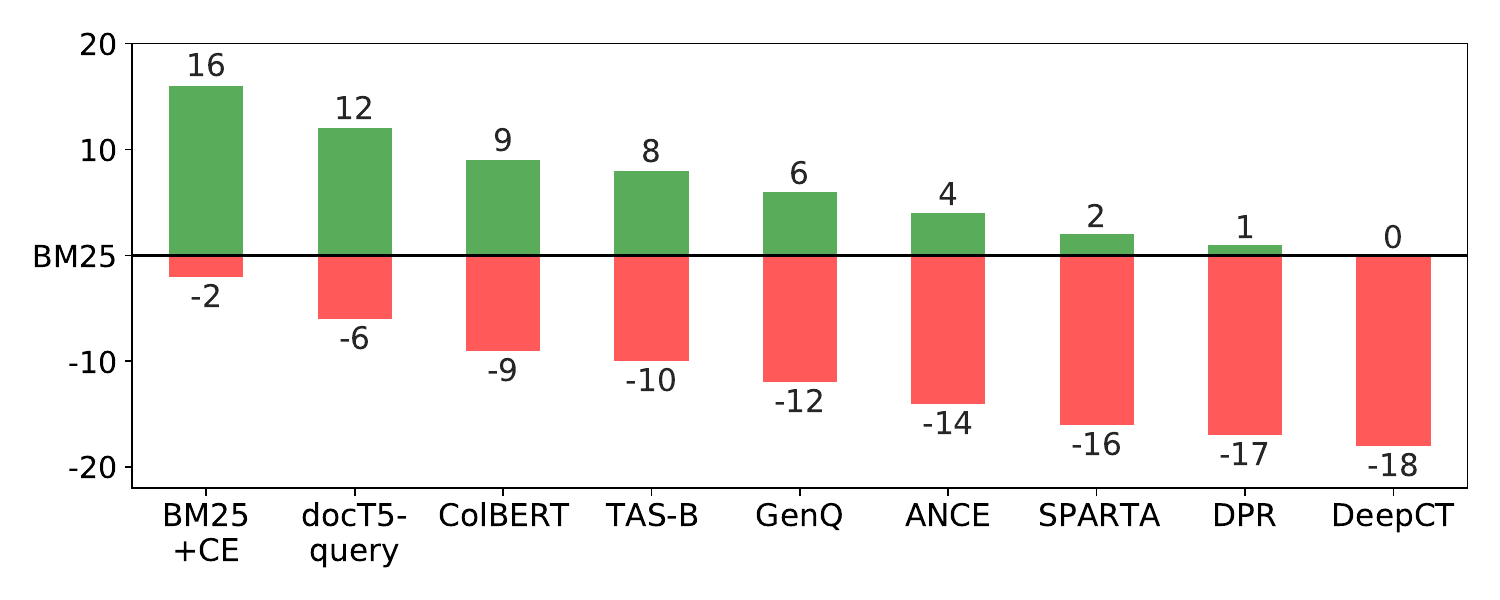}
}
{%
  \caption{Comparison of zero-shot neural retrieval performances with BM25. Re-ranking based models, i.e., BM25+CE and sparse model: docT5query outperform BM25 on more than half the \beir evaluation datasets. \vspace{-1mm} 
  }\label{fig:bm25-comparison}%
  }
  \ffigbox{%
\includegraphics[trim=0 0 0 0,clip,width=0.40\textwidth]{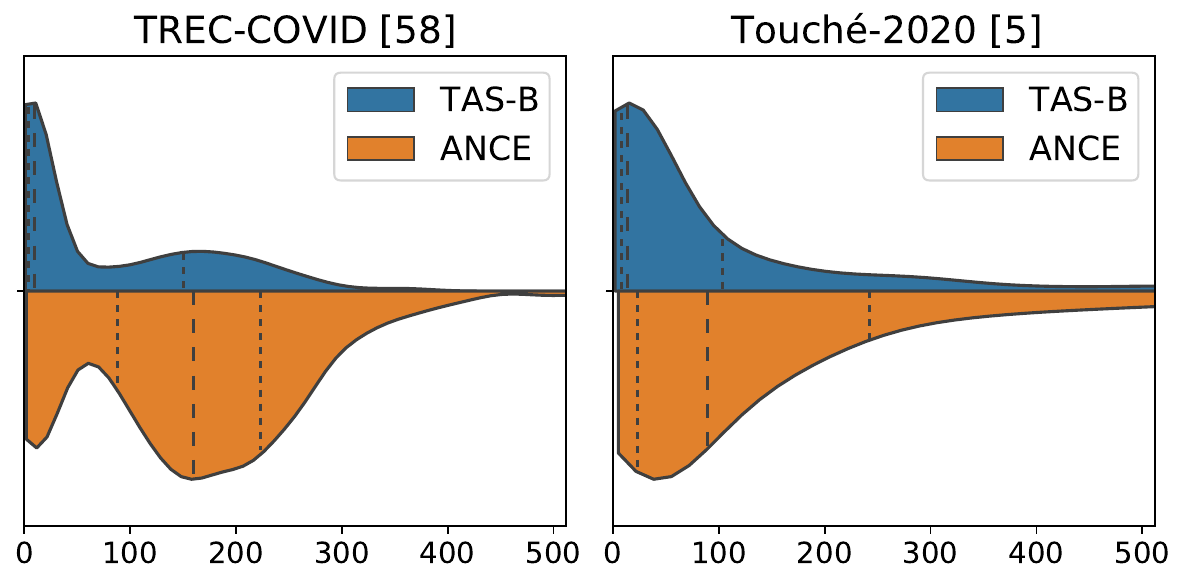}
}{%
  \caption{Distribution plots \cite{citeulike:4075875} for top-10 retrieved document lengths (in words)  using TAS-B (blue, top) or ANCE (orange, bottom). TAS-B has a preference towards shorter documents in \beir. \vspace{-1mm}}\label{fig:violin-plots}%
}
\end{floatrow}
\vspace*{-\baselineskip}
\end{figure}

In this section, we evaluate and analyze how retrieval models perform on the \beir benchmark. Table \ref{tab:results} reports the results of all evaluated systems on the selected benchmark datasets. As a baseline, we compare our retrieval systems against BM25. Figure \ref{fig:bm25-comparison} shows, on how many datasets a respective model is able to perform better or worse than BM25.

\begin{table*}[t!]
    \small
    \resizebox{\textwidth}{!}{\begin{tabular}{l | c | c c c | c c c c | c | c}
        \toprule
        \multicolumn{1}{l}{\textbf{Model ($\rightarrow$)}} &
        \multicolumn{1}{c}{Lexical}   &
        \multicolumn{3}{c}{Sparse}   &
        \multicolumn{4}{c}{Dense} &
        \multicolumn{1}{c}{Late-Interaction} &
        \multicolumn{1}{c}{Re-ranking} \\ 
        \cmidrule(lr){1-1}
        \cmidrule(lr){2-2}
        \cmidrule(lr){3-5}
        \cmidrule(lr){6-9}
        \cmidrule(lr){10-10}
        \cmidrule(lr){11-11}
        \multicolumn{1}{l}{\textbf{Dataset ($\downarrow$)}} &
        \multicolumn{1}{c}{\textbf{BM25}} &
        \multicolumn{1}{c}{\textbf{DeepCT}} &
        \multicolumn{1}{c}{\textbf{SPARTA}} &
        \multicolumn{1}{c}{\textbf{docT5query}} &
        \multicolumn{1}{c}{\textbf{DPR}} &
        \multicolumn{1}{c}{\textbf{ANCE}} &
        \multicolumn{1}{c}{\textbf{TAS-B}} &
        \multicolumn{1}{c}{\textbf{GenQ}} &
        \multicolumn{1}{c}{\textbf{ColBERT}} &
        \multicolumn{1}{c}{\textbf{BM25+CE}} \\
        \midrule

   MS MARCO & 0.228 & 0.296$^\ddagger$ & 0.351$^\ddagger$ & 0.338$^\ddagger$ & 0.177 & 0.388$^\ddagger$ & 0.408$^\ddagger$ & 0.408$^\ddagger$ & \underline{0.401}$^\ddagger$ & \textbf{0.413}$^\ddagger$ \\  \midrule \midrule
   
   TREC-COVID    & 0.656 & 0.406 & 0.538 & \underline{0.713} & 0.332& 0.654 & 0.481 & 0.619 & 0.677 & \textbf{0.757}\\
   BioASQ        & 0.465 & 0.407 & 0.351 & \textcolor{black}{0.431} & 0.127 & 0.306 & 0.383 & 0.398 & \underline{0.474} & \textbf{0.523} \\
   NFCorpus      & 0.325 & 0.283 & 0.301 & \underline{0.328} & 0.189 & 0.237 & 0.319 & 0.319 & 0.305 & \textbf{0.350} \\ \midrule
   NQ            & 0.329 & 0.188 & 0.398 & 0.399 & 0.474$^\ddagger$ & 0.446 & 0.463 & 0.358 & \underline{0.524} & \textbf{0.533} \\ 
   HotpotQA      & \underline{0.603} & 0.503 & 0.492 & 0.580 & 0.391 & 0.456 & 0.584 & 0.534 & 0.593 & \textbf{0.707} \\ 
   FiQA-2018     & 0.236 & 0.191 & 0.198 & 0.291 & 0.112 & 0.295 & 0.300 & 0.308 & \underline{0.317} & \textbf{0.347} \\ \midrule
   Signal-1M (RT)& \underline{0.330} & 0.269 & 0.252 & 0.307 & 0.155 & 0.249 & 0.289 & 0.281 & 0.274 & \textbf{0.338} \\ \midrule
   TREC-NEWS     & 0.398 & 0.220 & 0.258 & \underline{0.420} & 0.161 & 0.382 & 0.377 & 0.396 & 0.393 & \textbf{0.431} \\
   Robust04     & 0.408 & 0.287 & 0.276 & \underline{0.437} & 0.252 & 0.392 & 0.427 & 0.362 & 0.391 & \textbf{0.475} \\ \midrule
   ArguAna       & 0.315 & 0.309 & 0.279 & 0.349 & 0.175& 0.415 & \underline{0.429} & \textbf{0.493} & 0.233 & 0.311 \\  
   Touch\'e-2020 & \textbf{0.367} & 0.156 & 0.175 & \underline{0.347} & 0.131 & 0.240 & 0.162 & 0.182 & 0.202 & 0.271 \\ \midrule 
   CQADupStack   & 0.299 & 0.268 & 0.257 & 0.325 & 0.153 & 0.296 & 0.314 & 0.347 & \underline{0.350} & \textbf{0.370} \\
   Quora         & 0.789 & 0.691 & 0.630 & 0.802 & 0.248 & \underline{0.852} & 0.835 & 0.830 & \textbf{0.854} & 0.825 \\ \midrule
   DBPedia       & 0.313  & 0.177 & 0.314 & 0.331 & 0.263 & 0.281 & 0.384 & 0.328 & \underline{0.392} & \textbf{0.409} \\ \midrule
   SCIDOCS       & 0.158 & 0.124 & 0.126 & \underline{0.162} & 0.077 & 0.122 & 0.149 & 0.143 & 0.145 & \textbf{0.166} \\ \midrule
   FEVER         & 0.753 & 0.353 & 0.596 & \textcolor{black}{0.714} & 0.562& 0.669 & 0.700 & 0.669 & \underline{0.771} & \textbf{0.819} \\ 
   Climate-FEVER & 0.213 & 0.066 & 0.082 & 0.201 & 0.148 & 0.198 & \underline{0.228} & 0.175 & 0.184 & \textbf{0.253}\\ 
   SciFact       & 0.665 & 0.630 & 0.582 & \underline{0.675} & 0.318 & 0.507 & 0.643 & 0.644 & 0.671 & \textbf{0.688} \\ \midrule
   \multicolumn{2}{c|}{Avg. Performance vs. BM25} & \textbf{- 27.9\%} & \textbf{- 20.3\%} & \textbf{+ 1.6\%} & \textbf{- 47.7\%} & \textbf{- 7.4\%} & \textbf{ - 2.8\%} & \textbf{ - 3.6\%} & \textbf{+ 2.5\%} & \textbf{+ 11\%} \\ 
        \bottomrule
    \end{tabular}}
    \caption{In-domain and zero-shot performances on \beir benchmark. All scores denote \textbf{nDCG@10}. The best score on a given dataset is marked in \textbf{bold}, and the second best is \underline{underlined}. Corresponding Recall@100 performances can be found in Table \ref{tab:results-recall}. $\ddagger$ indicates the in-domain performances. \vspace{-5mm}}
    \label{tab:results}
\end{table*}

 \textbf{1. In-domain performance is not a good indicator for out-of-domain generalization.} We observe BM25 heavily underperforms neural approaches by 7-18 points on in-domain MS MARCO. However, \beir reveals it to be a strong baseline for generalization and generally outperforming many other, more complex approaches. This stresses the point, that retrieval methods must be evaluated on a broad range of datasets.
 
\textbf{2. Term-weighting fails, document expansion captures out-of-domain keyword vocabulary.}  DeepCT and SPARTA both use a transformer network to learn term weighting. While both methods perform well in-domain on MS MARCO, they completely fail to generalize well by under performing BM25 on nearly all datasets. In contrast, document expansion based docT5query is able to add new relevant keywords to a document and performs strong on the \beir datasets. It outperforms BM25 on 11/18 datasets while providing  a competitive performance on the remaining datasets.  

\textbf{3. Dense retrieval models with issues for out-of-distribution data.} Dense retrieval models (esp.\ ANCE and TAS-B), that map queries and documents independently to vector spaces, perform strongly on certain datasets, while on many other datasets perform significantly worse than BM25. For example, dense retrievers are observed to underperform on datasets with a large domain shift compared from what they have been trained on, like in BioASQ, or task-shifts like in Touch\'e-2020. DPR, the only non-MSMARCO trained dataset overall performs the worst in generalization on the benchmark.

\textbf{4. Re-ranking and Late-Interaction models generalize well to out-of-distribution data.} The cross-attentional re-ranking model (BM25+CE) performs the best and is able to outperform BM25 on almost all (16/18) datasets. It only fails on ArguAna and Touch\'e-2020, two retrieval tasks that are extremely different to the MS MARCO training dataset. The late-interaction model ColBERT computes token embeddings independently for the query and document, and scores (query, document)-pairs by a cross-attentional like MaxSim operation. It performs a bit weaker than the cross-attentional re-ranking model, but is still able to  outperform BM25 on 9/18 datasets. It appears that cross-attention and cross-attentional like operations are important for a good out-of-distribution generalization.

\textbf{5. Strong training losses for dense retrieval leads to better out-of-distribution performances.} TAS-B provides the best zero-shot generalization performance among its dense counterparts. It outperforms ANCE on 14/18 and DPR on 17/18 datasets respectively. We speculate that the reason lies in a strong training setup in combination of both in-domain batch negatives and Margin-MSE losses for the TAS-B model. This training loss function (with strong ensemble teachers in a Knowledge Distillation setup) shows strong generalization performances.

\textbf{6. TAS-B model prefers to retrieve documents with shorter lengths.} TAS-B underperforms ANCE on two datasets: TREC-COVID by 17.3 points and Touch\'e-2020 by 7.8 points. We observed that these  models retrieve documents with vastly different lengths as shown in Figure \ref{fig:violin-plots}. On TREC-COVID, TAS-B retrieves documents with a median length of mere 10 words versus ANCE with 160 words. Similarly on Touch\'e-2020, 14 words vs.\ 89 words with TAS-B and ANCE respectively. As discussed in Appendix \ref{sec:dense_retrieval_length_preference}, this preference for shorter or longer documents is due to the used loss function. 

\textbf{7. Does domain adaptation help improve generalization of dense-retrievers?} We evaluated GenQ, which further fine-tunes the TAS-B model on synthetic query data. It outperforms the TAS-B model on specialized domains like scientific publications, finance or StackExchange. On broader and more generic domains, like Wikipedia, it performs weaker than the original TAS-B model.

\vspace{-3mm}
\subsection{Efficiency: Retrieval Latency and Index Sizes}\label{sec:model-efficiency}
\vspace{-2mm}

Models need to potentially compare a single query against millions of documents at inference, hence, a high computational speed for retrieving results in real-time is desired. Besides speed, index sizes are vital and are often stored entirely in memory. We randomly sample 1 million documents from DBPedia \cite{Hasibi:2017:DVT} and evaluate latency. For dense models, we use exact search, while for ColBERT we follow the original setup \cite{10.1145/3397271.3401075} and use approximate nearest neighbor search. Performances on CPU were measured with an 8 core Intel Xeon Platinum 8168 CPU~@~2.70GHz and on GPU using a single Nvidia Tesla V100, CUDA 11.0. 

\begin{wraptable}{r}{0.45\textwidth}
  \vspace{-7mm}
  \small
  \ra{1.1}
  \resizebox{\textwidth}{!}{\begin{tabular}{c l | c | c | c |  c}
        \toprule
        \multicolumn{3}{c}{DBPedia \cite{Hasibi:2017:DVT} (1 Million)} &
        \multicolumn{2}{|c}{Retrieval Latency} &
        \multicolumn{1}{c}{Index} \\
        \cmidrule(lr){1-3}
        \cmidrule(lr){4-5}
        \cmidrule(lr){6-6}
        \multicolumn{1}{c}{Rank} &
        \multicolumn{1}{l}{Model} &
        \multicolumn{1}{|c|}{Dim.} &
        \multicolumn{1}{c}{GPU} & 
        \multicolumn{1}{c}{CPU} &
        \multicolumn{1}{c}{Size} \\ \midrule
        (1) & BM25+CE      & --  & 450ms & 6100ms & 0.4GB \\
        (2) & ColBERT      & 128 & 350ms& -- & 20GB \\
        (3) & docT5query   & --  &   -- & 30ms& 0.4GB \\
        (4) & BM25         & --  &   -- & 20ms& 0.4GB \\
        (5) & TAS-B        & 768 & 14ms & 125ms& 3GB \\
        (6) & GenQ         & 768 & 14ms & 125ms& 3GB \\
        (7) & ANCE         & 768 & 20ms & 275ms& 3GB \\
        (8) & SPARTA       & 2000&  --  & 20ms & 12GB \\
        (9) & DeepCT       & --  &   -- & 25ms& 0.4GB \\
        (10) & DPR         & 768 & 19ms & 230ms & 3GB \\
        \bottomrule
    \end{tabular}}
    \caption{Estimated average retrieval latency and index sizes for a single query in DBPedia \cite{Hasibi:2017:DVT}. Ranked from best to worst on zero-shot \beir. Lower the latency or memory is desired. \vspace{-3mm}}
\end{wraptable}

\textbf{Tradeoff between performance and retrieval latency} \quad The best out-of-distribution generalization performances by re-ranking top-100 BM25 documents and with late-interaction models come at the cost of high latency (> 350 ms), being slowest at inference. In contrast, dense retrievers are 20-30x faster (< 20ms) compared to the re-ranking models and follow a low-latency pattern. On CPU, the sparse models  dominate in terms of speed (20-25ms). 

\textbf{Tradeoff between performance and index sizes} \quad Lexical, re-ranking and dense methods have the smallest index sizes (< 3GB) to store 1M documents from DBPedia. SPARTA requires the second largest index to store a 30k dim sparse vector while ColBERT requires the largest index as it stores multiple 128 dim dense vectors for a single document. Index sizes are especially relevant when document sizes scale higher: ColBERT requires \char`\~900GB to store the BioASQ (\char`\~15M documents) index, whereas BM25 only requires 18GB. 

\vspace{-3mm}
\section{Impact of Annotation Selection Bias}\label{sec:biases-analysis}
\vspace{-2mm}

Creating a perfectly unbiased evaluation dataset for retrieval is inherently complex and is subject to multiple biases induced by the: (\emph{i}) annotation guidelines, (\emph{ii}) annotation setup, and by the (\emph{iii}) human annotators. Further, it is impossible to manually annotate the relevance for all (query, document)-pairs. Instead, existing retrieval methods are used to get a pool of candidate documents which are then marked for their relevance. All other unseen documents are assumed to be irrelevant.
This is a source for \textit{selection bias} \cite{selection_bias}: A new retrieval system might retrieve vastly different results than the system used for the annotation. These hits are automatically assumed to be irrelevant.

Many \beir datasets are found to be subject to a lexical bias, i.e.\ a lexical based retrieval system like TF-IDF or BM25 has been used to retrieve the candidates for annotation. For example, in BioASQ, candidates have been retrieved for annotation via term-matching with boosting tags \cite{tsatsaronis2015overview}. Creation of Signal-1M (RT) involved retrieving tweets for a query with 7 out of these 8 techniques relying upon lexical term-matching signals \cite{Signal1MRelatedTweetsRetrieval2018}. Such a lexical bias disfavours approaches that don't rely on lexical matching, like dense retrieval methods, as retrieved hits without lexical overlap are automatically assumed to be irrelevant, even though the hits might be relevant for a query.

In order to study the impact of this particular type of bias, we conducted a study on the recent TREC-COVID dataset. TREC-COVID used a pooling method \cite{10.1145/2911451.2911473, lipani2016curious} to reduce the impact of the aforementioned bias: The annotation set was constructed by using the search results from the various systems participating in the challenge.  Table \ref{tab:hole@k} shows the Hole@10 rate \cite{xiong2020approximate} for the tested systems, i.e., how many top-10 hits is each system retrieving that have not been seen by annotators.

\begin{table*}[t!]
    \small
    \resizebox{\textwidth}{!}{\begin{tabular}{l | c | c c c | c c c | c | c}
        \toprule
        \multicolumn{1}{l}{\textbf{Model ($\rightarrow$)}} &
        \multicolumn{1}{|c}{\textbf{BM25}} &
        \multicolumn{1}{|c}{\textbf{DeepCT}} &
        \multicolumn{1}{c}{\textbf{SPARTA}} &
        \multicolumn{1}{c}{\textbf{docT5query}} &
        \multicolumn{1}{|c}{\textbf{DPR}} &
        \multicolumn{1}{c}{\textbf{ANCE}} &
        \multicolumn{1}{c}{\textbf{TAS-B}} &
        \multicolumn{1}{|c|}{\textbf{ColBERT}} &
        \multicolumn{1}{c}{\textbf{BM25+CE}} \\
        \midrule
  Hole@10 (in \%)   & 6.4\% & 19.4\% & 12.4\% & 2.8\% & 30.6\% & 14.4\% & 31.8\% & 12.4\% & 1.6\% \\ \midrule 
  \multicolumn{10}{c}{nDCG@10 performances before and after manual annotation on TREC-COVID \cite{10.1145/3451964.3451965}} \\ \midrule
  Original (w/ holes) & 0.656 & 0.406 & 0.538 & \underline{0.713} & 0.332 & 0.654 & 0.481 & 0.677 & \textbf{0.757}\\ \midrule
  Annotated (w/o holes) & 0.668 & 0.472 & 0.624 & 0.714 & 0.445 & \underline{0.735} & 0.555 & \underline{0.735} & \textbf{0.760} \\
   
    \bottomrule
    \end{tabular}}
    \caption{Hole@10 analysis on TREC-COVID. Annotated scores show improvement in performances after removing holes@10 (documents in top-10 hits unseen by annotators) across each model. \vspace{-3mm}}
    \label{tab:hole@k}
\end{table*}

The results reveal large differences between approaches: Lexical approaches like BM25 and docT5query have a rather low Hole@10 value of 6.4\% and 2.8\%, indicating that the annotation pool contained the top-hits from lexical retrieval systems. In contrast, dense retrieval systems like ANCE and TAS-B have a much higher Hole@10 of 14.4\% and 31.8\%, indicating that a large fraction of hits found by these systems have not been judged by annotators.
Next, we manually added for all systems, the missing annotation (or holes) following the original annotation guidelines. During annotation, we were unaware of the system who retrieved the missing annotation to avoid a preference bias. In total, we annotated 980 query-document pairs in TREC-COVID. We then re-computed nDCG@10 for all systems with this additional annotations.

As shown in Table \ref{tab:hole@k}, we observe that lexical approaches improves only slightly, e.g.\ for docT5query just from 0.713 to 0.714 after adding the missing relevance judgements. In contrast, for the dense retrieval system ANCE, the performance improves from 0.654 (slightly below BM25) to 0.735, which is 6.7 points above the BM25 performance. Similar improvements are noticed in ColBERT (5.8 points). Even though many systems contributed to the TREC-COVID annotation pool, the annotation pool is still biased towards lexical approaches.

\vspace{-3mm}
\section{Conclusions and Future Work}
\vspace{-2mm}

In this work, we presented \beir: a heterogeneous benchmark for information retrieval. We provided a broader selection of target tasks ranging from narrow expert domains to open domain datasets. We included nine different retrieval tasks spanning 18 diverse datasets.  

By open-sourcing \beir, with a standardized data format and easy-to-adapt code examples for many different retrieval strategies, we take an important steps towards a unified benchmark to evaluate the zero-shot capabilities of retrieval systems. It hopefully steers innovation towards more robust retrieval systems and to new insights which retrieval architectures perform well across tasks and domains. 

We studied the effectiveness of ten different retrieval models and demonstrate, that in-domain performance cannot predict how well an approach will generalize in a zero-shot setup. Many approaches that outperform BM25 on an in-domain evaluation, perform poorly on the \beir datasets.  Cross-attentional re-ranking, late-interaction ColBERT, and the document expansion technique docT5query performed overall well across the evaluated tasks. 

Our study on annotation selection bias highlights the challenge of evaluating new models on existing datasets: Even though TREC-COVID is based on the predictions from many systems, contributed by a diverse set of teams, we found largely different \textit{Hole@10} rates for the tested systems, negatively affecting non-lexical approaches. Better datasets, that use diverse pooling strategies, are needed for a fair evaluation of retrieval approaches. By integrate a large number of diverse retrieval systems into BEIR, creating such diverse pools becomes significantly simplified.

\clearpage

\bibliography{custom}
\bibliographystyle{acl_natbib}

\clearpage

\section*{Checklist}


\begin{enumerate}

\item For all authors...
\begin{enumerate}
  \item Do the main claims made in the abstract and introduction accurately reflect the paper's contributions and scope?
    \answerYes{}
  \item Did you describe the limitations of your work?
    \answerYes{See Appendix~\ref{sec:limitations}.}
  \item Did you discuss any potential negative societal impacts of your work?
    \answerNo{}
  \item Have you read the ethics review guidelines and ensured that your paper conforms to them?
    \answerYes{}
\end{enumerate}

\item If you are including theoretical results...
\begin{enumerate}
  \item Did you state the full set of assumptions of all theoretical results?
    \answerNA{}
	\item Did you include complete proofs of all theoretical results?
    \answerNA{}
\end{enumerate}

\item If you ran experiments (e.g. for benchmarks)...
\begin{enumerate}
  \item Did you include the code, data, and instructions needed to reproduce the main experimental results (either in the supplemental material or as a URL)?
    \answerYes{URL mentioned in Abstract.}
  \item Did you specify all the training details (e.g., data splits, hyperparameters, how they were chosen)?
    \answerYes{All results can be reproduced by the code in our repository.}
	\item Did you report error bars (e.g., with respect to the random seed after running experiments multiple times)?
    \answerNo{We evaluate existing available pre-trained models that often come without suitable training code. Hence, in many cases, re-training the model is not feasible.}
	\item Did you include the total amount of compute and the type of resources used (e.g., type of GPUs, internal cluster, or cloud provider)?
    \answerNo{We include the type of GPU and CPU resources we used, but not the total amount of compute that was used.}
\end{enumerate}

\item If you are using existing assets (e.g., code, data, models) or curating/releasing new assets...
\begin{enumerate}
  \item If your work uses existing assets, did you cite the creators?
    \answerYes{Original papers are cited (if available), Table \ref{tab:dataset_links} contains the original website links for the used datasets.}
  \item Did you mention the license of the assets?
    \answerYes{See Appendix \ref{sec:dataset_licenses}.}
  \item Did you include any new assets either in the supplemental material or as a URL?
    \answerYes{No supplemental material attached to this submission. Further supplemental material can be found in our repository mentioned in the URL.}
  \item Did you discuss whether and how consent was obtained from people whose data you're using/curating?
    \answerNA{Used datasets provide a specific dataset license, which we follow.}
  \item Did you discuss whether the data you are using/curating contains personally identifiable information or offensive content?
    \answerNo{We re-use existing datasets, which most are freely available. Most datasets are from less sensitive sources, like Wikipedia or scientific publications, where don't expect  personally identifiable information. Checking for offensive content in more than 50 million documents is difficult and removing it would alter the underlying dataset.}
\end{enumerate}

\item If you used crowdsourcing or conducted research with human subjects...
\begin{enumerate}
  \item Did you include the full text of instructions given to participants and screenshots, if applicable?
    \answerNA{We ourselves performed annotation on the TREC-COVID dataset, where we followed the instructions from the original task website.}
  \item Did you describe any potential participant risks, with links to Institutional Review Board (IRB) approvals, if applicable?
    \answerNA{}
  \item Did you include the estimated hourly wage paid to participants and the total amount spent on participant compensation?
    \answerNA{Annotations were done by the authors of the paper.}
\end{enumerate}

\end{enumerate}

\appendix

\section{Complementing Information}
We provide the following additional sections in detail and information that complement discussions in the main paper: 

\begin{enumerate}
    \item[$\bullet$] Limitations of the \beir benchmark in Appendix \ref{sec:limitations}.
    \item[$\bullet$] Training and in-domain evaluation task details in Appendix \ref{sec:training_dataset}.
    \item[$\bullet$] Description of all zero-shot tasks and datasets used in \beir in Appendix \ref{sec:datasets}.
    \item[$\bullet$] Details of dataset licenses in Appendix \ref{sec:dataset_licenses}.
    \item[$\bullet$] Overview of the weighted jaccard similarity metric in Appendix \ref{sec:weighted_jaccard_similarity}.
    \item[$\bullet$] Overview of the capped recall at k metric in Appendix \ref{sec:capped_recall_score}.
    \item[$\bullet$] Length preference for dense retrieval system in Appendix \ref{sec:dense_retrieval_length_preference}.
\end{enumerate}

\vspace{-2mm}
\section{Limitations of the BEIR Benchmark}\label{sec:limitations}
\vspace{-2mm}

Even though we cover a wide range of tasks and domains in \beir, no benchmark is perfect and has its limitations. Making those explicit is a critical point in understanding the results on the benchmark and, for future work, to improve up-on the benchmark. 

\textbf{1. Multilingual Tasks:} Although we aim for a diverse retrieval evaluation benchmark, due to the limited availability of multilingual retrieval datasets, all datasets covered in the \beir benchmark are currently English. It is worthwhile to add more multilingual datasets \cite{asai-etal-2021-xor, zhang2021mr} (in consideration of the selection criteria) as a next step for the benchmark. Future work could include multi- and cross-lingual tasks and models.

\textbf{2. Long Document Retrieval:} Most of our tasks have average document lengths up-to a few hundred words roughly equivalent to a few paragraphs. Including tasks that require the retrieval of longer documents would be highly relevant. However, as transformer-based approaches often have a length limit of 512 word pieces, a fundamental different setup would be required to compare approaches.

\textbf{3. Multi-factor Search:} Until now, we focused on pure textual search in \beir. In many real-world applications, further signals are used to estimate the relevancy of documents, such as PageRank \cite{ilprints422}, recency \cite{10.1145/1772690.1772725}, authority score \cite{10.1145/324133.324140} or user-interactions such as click-through rates \cite{10.1145/1458082.1458092}. The integration of such signals in the tested approaches is often not straight-forward and is an interesting direction for research.

\textbf{4. Multi-field Retrieval:} Retrieval can often be performed over multiple fields. For example, for scientific publication we have the title, the abstract, the document body, the authors list, and the journal name. So far we focused only on datasets that have one or two fields.

\textbf{5. Task-specific Models:} In our benchmark, we focus on evaluating models that are able to generalize well for a broad range of retrieval tasks. Naturally in real-world, for some few tasks or domains, specialized models are available which can easily outperform generic models as they focus and perform well on a single task, lets say on question-answering. Such task-specific models do not necessarily need to generalize across all diverse tasks.

\vspace{-3mm}
\section{Training and In-domain Evaluation}\label{sec:training_dataset}
\vspace{-2mm}

We use the MS MARCO Passage Ranking dataset \cite{nguyen2016ms}, which contains 8.8M Passages and an official training set of 532,761 query-passage pairs for fine-tuning for a majority of retrievers. The dataset contains queries from Bing search logs with one text passage from various web sources annotated as relevant. We find the dataset useful for training, in terms of covering a wide variety of topics and providing the highest number of training pairs. It has been extensively explored and used for fine-tuning dense retrievers in recent works \cite{nogueira2020passage,gao2020complementing,ding2020rocketqa}. We use the official MS MARCO development set for our in-domain evaluation
which has been widely used in prior research \cite{nogueira2020passage, gao2020complementing,ding2020rocketqa}. It has 6,980 queries. Most of the queries have only 1 document judged relevant; the labels are binary.

\vspace{-2mm}
\section{Zero-shot Evaluation Tasks}\label{sec:datasets}
\vspace{-2mm}

Following the selection criteria mentioned in Section \ref{sec_beir_benchmark}, we include 18 evaluation datasets that span across 9 heterogeneous tasks. Each dataset mentioned below contains a document corpus denoted by $\mathbf{T}$ and test queries for evaluation denoted by $\mathbf{Q}$. We additionally provide dataset website links in Table \ref{tab:dataset_links} and intuitive examples in Table \ref{tab:examples}. We now describe each task and dataset included in the \beir benchmark below:

\vspace{-2mm}
\subsection{Bio-Medical Information Retrieval}
\vspace{-2mm}
Bio-medical information retrieval is the task of searching relevant scientific documents such as research papers or blogs for a given scientific query in the biomedical domain \cite{jiang2007empirical}. We consider a scientific query as \textit{input} and retrieve bio-medical documents as \textit{output}.

\textbf{TREC-COVID} \cite{10.1145/3451964.3451965} is an ad-hoc search challenge based on the CORD-19 dataset containing scientific articles related to the COVID-19 pandemic \cite{wang2020cord19}. We include the July 16, 2020 version of CORD-19 dataset as corpus $\mathbf{T}$ and use the final cumulative judgements with query descriptions from the original task as queries $\mathbf{Q}$.

\textbf{NFCorpus} \cite{boteva2016} contains natural language queries harvested from NutritionFacts (NF). We use the original splits provided alongside all content sources from NF (videos, blogs, and Q\&A posts) as queries $\mathbf{Q}$ and annotated medical documents from PubMed as corpus $\mathbf{T}$.

\textbf{BioASQ} \cite{tsatsaronis2015overview} Task 8b is a biomedical semantic question answering challenge. We use the original train and test splits provided in Task 8b as queries $\mathbf{Q}$ and collect around 15M articles from PubMed provided in Task 8a as our corpus $\mathbf{T}$. 

\vspace{-1mm}
\subsection{Open-domain Question Answering (QA)}
\vspace{-2mm}

Retrieval in open domain question answering \cite{chen-etal-2017-reading} is the task of retrieving the correct answer for a question, without a predefined location for the answer. In open-domain tasks, model must retrieve over an entire knowledge source (such as Wikipedia). We consider the question as \textit{input} and the passage containing the answer as \textit{output}.

\textbf{Natural Questions} \cite{47761} contains Google search queries and documents with paragraphs and answer spans within Wikipedia articles. We did not use the NQ version from ReQA \cite{ahmad-etal-2019-reqa} as it focused on queries having a short answer. As a result, we parsed the HTML of the original NQ dataset and include more complex development queries that often require a longer passage as answer compared to ReQA. We filtered out queries without an answer, or having a table as an answer, or with conflicting Wikipedia pages. We retain 2,681,468 passages as our corpus $\mathbf{T}$ and 3452 test queries $\mathbf{Q}$ from the original dataset.

\textbf{HotpotQA} \cite{yang-etal-2018-hotpotqa} contains multi-hop like questions which require reasoning over multiple paragraphs to find the correct answer. We include the original full-wiki task setting: utilizing processed Wikipedia passages as corpus $\mathbf{T}$. We held out randomly sampled 5447 queries from training as our dev split. We use the original (paper) task's development split as our test split $\mathbf{Q}$.

 \textbf{FiQA-2018} \cite{10.1145/3184558.3192301} Task 2 consists of opinion-based question-answering. We include financial data by crawling StackExchange posts under the Investment topic from 2009-2017 as our corpus $\mathbf{T}$. We randomly sample out 500 and 648 queries $\mathbf{Q}$ from the original training split as dev and test splits.

\vspace{-2mm}
\subsection{Tweet Retrieval}
\vspace{-1mm}
Twitter is a popular micro-blogging website on which people post real-time messages (i.e.\ tweets) about their opinions on a variety of topics and discuss current issues. We consider a news headline as \textit{input} and retrieve relevant tweets as \textit{output}.

\textbf{Signal-1M Related Tweets} \cite{Signal1MRelatedTweetsRetrieval2018} task retrieves relevant tweets for a given news article title. The Related Tweets task provides news articles from the Signal-1M dataset \cite{Signal1M2016} which we use as queries $\mathbf{Q}$. We construct our twitter corpus $\mathbf{T}$ by manually scraping tweets from the provided tweet-ids in the relevancy judgements using Python package: Tweepy (https://www.tweepy.org).

\vspace{-2mm}
\subsection{News Retrieval}
\vspace{-1mm}

\textbf{TREC-NEWS} \cite{soboroff2019trec} 2019 track involves background linking: Given a news headline, we retrieve relevant news articles that provide important context or background information. We include the original shared task query description (single sentence) as our test queries $\mathbf{Q}$ and the TREC Washington Post as our corpus $\mathbf{T}$. For simplicity, we convert the original exponential gain relevant judgements to linear labels. 

\textbf{Robust04} \cite{96071} provides a robust dataset focusing on evaluating on poorly performing topics. We include the original shared task query description (single sentence) as our test queries $\mathbf{Q}$ and the complete TREC disks 4 and 5 documents as our corpus $\mathbf{T}$.

\vspace{-2mm}
\subsection{Argument Retrieval}
\vspace{-1mm}
Argument retrieval is the task of ranking argumentative texts in a collection of focused arguments (\textit{output}) in order of their relevance to a textual query (\textit{input}) on different topics.

\textbf{ArguAna Counterargs Corpus} \cite{wachsmuth:2018a} involves the task of retrieval of the best counterargument to an argument. We include pairs of arguments and counterarguments scraped from the online debate portal as corpus $\mathbf{T}$. We consider the arguments present in the original test split as our queries $\mathbf{Q}$.

\textbf{Touch\'e-2020} \cite{stein:2020v} Task 1 is a conversational argument retrieval task. We use the conclusion as title and premise for arguments present in args.me \cite{stein:2017r} as corpus $\mathbf{T}$. We include the shared Touch\'e-2020 task data as our test queries $\mathbf{Q}$. The original relevance judgements (qrels) file also included negative judgements (-2) for non-arguments present within the corpus, but for simplicity we substitute them as zero.  

\vspace{-2mm}
\subsection{Duplicate Question Retrieval}
\vspace{-1mm}

Duplicate question retrieval is the task of identifying duplicate questions asked in community question answering (cQA) forums. A given query is the \textit{input} and the duplicate questions are the \textit{output}.

\textbf{CQADupStack} \cite{hoogeveen2015cqadupstack} is a popular dataset for research in community question-answering (cQA). The corpus $\mathbf{T}$ comprises of queries from 12 different StackExchange subforums: Android, English, Gaming, Gis, Mathematica, Physics, Programmers, Stats, Tex, Unix, Webmasters and Wordpress. We utilize the original test split for our queries $\mathbf{Q}$, and the task involves retrieving duplicate query (title + body) for an input query title. We evaluate each StackExchange subforum separately and report the overall mean scores for all tasks in \beir.

\textbf{Quora} Duplicate Questions dataset identifies whether two questions are duplicates. Quora originally released containing 404,290 question pairs. We add transitive closures to the original dataset. Further, we split it into train, dev, and test sets with a ratio of about 85\%, 5\% and 10\% of the original pairs. We remove all overlaps between the splits and ensure that a question in one split of the dataset does not appear in any other split to mitigate the transductive classification problem \cite{10.1007/978-3-642-15880-3_42}. We achieve 522,931 unique queries as our corpus $\mathbf{T}$ and 5,000 dev and 10,000 test queries $\mathbf{Q}$ respectively.

\vspace{-2mm}
\subsection{Entity Retrieval}
\vspace{-1mm}
Entity retrieval involves retrieving unique Wikipedia pages to entities mentioned in the query. This is crucial for tasks involving Entity Linking (EL). The entity-bearing query is the \textit{input} and the entity abstract and title are retrieved as \textit{output}.

\textbf{DBPedia-Entity-v2} \cite{Hasibi:2017:DVT} is an established entity retrieval dataset. It contains a set of heterogeneous entity-bearing queries $\mathbf{Q}$ containing named entities, IR style keywords, and natural language queries. The task involves retrieving entities from the English part of DBpedia corpus $\mathbf{T}$ from October 2015. We randomly sample out 67 queries from the test split as our dev set.

\vspace{-2mm}
\subsection{Citation Prediction}
\vspace{-1mm}
Citations are a key signal of relatedness between scientific papers \cite{cohan-etal-2020-specter}. In this task, the model attempts to retrieve cited papers (\textit{output}) for a given paper title as \textit{input}.

\textbf{SCIDOCS} \cite{cohan-etal-2020-specter} contains a corpus $\mathbf{T}$ of 30K held-out pool of scientific papers. We consider the direct-citations (1 out of 7 tasks mentioned in the original paper) as the best suited task for retrieval evaluation in \beir. The task includes 1k papers as queries $\mathbf{Q}$ with 5 relevant papers and 25 (randomly selected) uncited papers for each query. 

\vspace{-1mm}
\subsection{Fact Checking}
\vspace{-1mm}
Fact checking verifies a claim against a big collection of evidence \cite{thorne-etal-2018-fever}. The task requires knowledge about the claim and reasoning over multiple documents. We consider a sentence-level claim as \textit{input} and the relevant document passage verifying the claim as \textit{output}.

\textbf{FEVER} \cite{thorne-etal-2018-fever} The Fact Extraction and VERification dataset is collected to facilitate the automatic fact checking. We utilize the original paper splits as queries $\mathbf{Q}$ and retrieve evidences from the pre-processed Wikipedia Abstracts (June 2017 dump) as our corpus $\mathbf{T}$.

\textbf{Climate-FEVER} \cite{diggelmann2020climatefever} is a dataset for verification of real-world climate claims. We include the original dataset claims as queries $\mathbf{Q}$ and retrieve evidences from the same FEVER Wiki corpus $\mathbf{T}$. We manually included few Wikipedia articles (25) missing from our corpus, but present within our relevance judgements.

\textbf{SciFact} \cite{wadden-etal-2020-fact} verifies scientific claims using evidence from the research literature containing scientific paper abstracts. We use the original publicly available dev split from the task containing 300 queries as our test queries $\mathbf{Q}$, and include all documents from the original dataset as our corpus $\mathbf{T}$.

\vspace{-2mm}
\section{Dataset Licenses}\label{sec:dataset_licenses}
\vspace{-2mm}

The authors of 4 out of the 19 datasets in the \beir benchmark (NFCorpus, FiQA-2018, Quora, Climate-Fever) do not report the dataset license in the paper or a repository; We overview the rest:

\begin{enumerate}
    \item[$\bullet$] MSMARCO: Provided under ``MIT License'' for non-commercial research purposes.
    \item[$\bullet$] FEVER, NQ, DBPedia, Signal-1M: All provided under CC BY-SA 3.0 license.
    \item[$\bullet$] TREC-NEWS, Robust04, BioASQ: Data collection archives are under \textbf{Copyright}.
    \item[$\bullet$] ArguAna, Touch\'e-2020: Provided under CC BY 4.0 license.
    \item[$\bullet$] CQADupStack: Provided under Apache License 2.0 license.
    \item[$\bullet$] SciFact: Provided under the CC BY-NC 2.0 license.
    \item[$\bullet$] SCIDOCS: Provided under the GNU General Public License v3.0 license.
    \item[$\bullet$] HotpotQA: Provided under the CC BY-SA 4.0 license.
    \item[$\bullet$] TREC-COVID: Provided under the ``Dataset License Agreement''.
    
\end{enumerate}

\section{Weighted Jaccard Similarity}\label{sec:weighted_jaccard_similarity}
\vspace{-2mm}

The weighted Jaccard similarity $J(S,T)$ \cite{ioffe2010improved} is intuitively calculated as the unique word overlap for all words present in both the datasets. More formally, the normalized frequency for an unique word $k$ in a dataset is calculated as the frequency of word $k$ divided over the sum of frequencies of all words in the dataset. 

${S_k}$ is the normalized frequency of word $k$ in the source dataset $S$ and ${T_k}$ for the target dataset $T$ respectively. The weighted Jaccard similarity between $S$ and $T$ is defined as:

\[ J(S, T) = \frac{\sum_{k}\min(S_k, T_k)}{\sum_{k}\max(S_k, T_k)}\] 

where the sum is over all unique words $k$ present in datasets $S$ and $T$.

\section{Capped Recall@k Score}\label{sec:capped_recall_score}
\vspace{-2mm}

Recall at $k$ is calculated as the fraction of the relevant documents that are successfully retrieved within the top $k$ extracted documents. More formally, the $R@k$ score is calculated as:

\[ R@k = \frac{1}{|Q|} \sum_{i=1}^{|Q|}\frac{|\max_k(A_i) \cap A_i^\star|}{|A_i^\star|} \]

where $Q$ is the set of queries, $A_i^\star$ is the set of relevant documents for the $i$th query, and $A_i$ is a scored list of documents provided by the model, from which top $k$ are extracted. 

However measuring recall can be counterintuitive, if a high number of relevant documents ($>k$) are present within a dataset. For example, consider a hypothetical dataset with 500 relevant documents for a query. Retrieving all relevant documents would produce a maximum $R@100$ score = 0.2, which is quite low and unintuitive. To avoid this we cap the recall score ($R\_cap@k$) at k for datasets if the number of relevant documents for a query greater than k. It is defined as: 

\[ R\_cap@k = \frac{1}{|Q|} \sum_{i=1}^{|Q|}\frac{|\max_k(A_i) \cap A_i^\star|}{\min(k,|A_i^\star|)} \] 

where the only difference lies within the denominator where we compute the minimum of k and $|A_i^\star|$, instead of $|A_i^\star|$ present in the original recall.

\section{Document Length Preference for Dense Retrieval System} \label{sec:dense_retrieval_length_preference} 

As we show in Figure \ref{fig:violin-plots}, TAS-B prefers retrieval of shorter documents, and in comparison, ANCE retrieves longer documents. The difference is especially extreme for the TREC-COVID dataset: TAS-B retrieves lots of top hit documents containing only a title and an empty abstract, while ANCE retrieves top hit documents with a non-empty abstract.

Identifying the source for this contrasting behaviour is difficult, as TAS-B and ANCE use different models (DistilBERT vs. RoBERTa-base), a different loss function (InfoNCE \cite{oord2019representation} vs. Margin-MSE \cite{hofstatter2021improving} with in-batch negatives), and different hard negative mining strategies. Hence, we decided to harmonize the training setup and to alter the training by just one aspect: The similarity function.

Dense models require a similarity function to retrieve relevant documents for a given query within an embedding space. This similarity function is also used during training dense models with the InfoNCE \cite{oord2019representation} loss:

$$\mathcal{L}_q = -\log \frac{\exp(\tau \cdot \text{sim}(q, d_+))}{\sum_{i=0}^n \exp(\tau \cdot \text{sim}(q, d_i))}$$

using $n$ in-batch negatives for each query $q$ and a scaling factor $\tau$. where $d_+$ denotes the relevant (positive) document for query $q$. Commonly used similarity functions ($\text{sim}(q, d)$) are cosine-similarity or dot-product. 

We trained two distilbert-base-uncased models with an identical training setup on MS MARCO (identical training parameters) and only changed the similarity function from cosine-similarity to dot-product. As shown in Table \ref{tab:trec-covid}, we observe significant performance differences for some \beir datasets. For TREC-COVID, the dot-product model achieves the biggest improvement with 15.3 points, while for a majority on other datasets, it performs worse than the cosine-similarity model.

We observe that these (nearly) identical models retrieve documents with vastly different lengths as shown in the violin plots in Table \ref{tab:trec-covid}. For all datasets, we find the cosine-similarity model to prefer shorter documents over longer ones. This is especially severe for TREC-COVID: a large fraction of the scientific papers (approx.~42k out of 171k) consist only of publication titles without an abstract. The cosine-similarity model prefers retrieving these documents. In contrast, the dot-product model primarily retrieves longer documents, i.e., publications with an abstract. Cosine-similarity uses vectors of unit length, thereby having no notion of the encoded text length. In contrast, for dot-product, longer documents result in vectors with higher magnitudes which can yield higher similarity scores for a query.

\begin{figure}[htb!]
\centering
\begin{center}
    \includegraphics[trim=7 7 10 10,clip,width=0.8\textwidth]{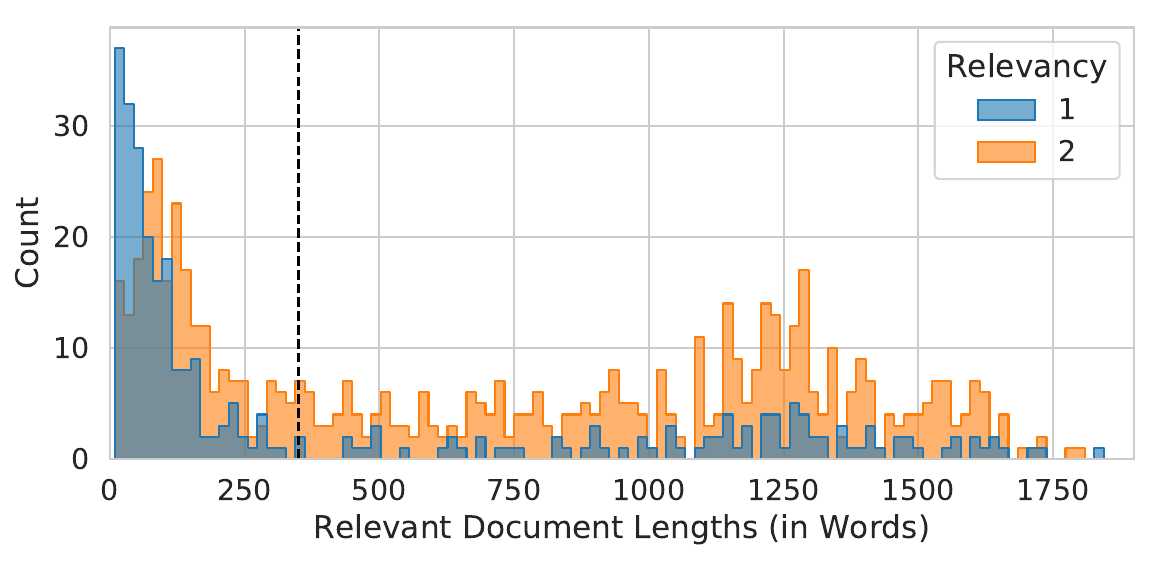}
    \caption{Annotated original relevant document lengths (in words) for Touch\'e-2020 \cite{stein:2020v}. Majority of the relevant documents (score = 2) on average in the original dataset are longer. Many shorter documents are annotated as less relevant (score = 1).}
    \label{fig:touche-task}
    \vspace*{-\baselineskip}
\end{center}
\end{figure}

Further, as we observe in Figure \ref{fig:touche-task}, relevance judgement scores are not uniformly distributed over document lengths: for some datasets, longer documents are annotated with higher relevancy scores, while in others, shorter documents are. This can be either due to the annotation process, e.g., the candidate selection method prefers short or long documents, or due to the task itself, where shorter or longer documents could be more relevant to the user information need. Hence, it can be more advantageous to train a model with either cosine-similarity or dot-product depending upon the nature and needs of the specific task.

\clearpage


\begin{table*}[t]
    \small
    \ra{1.2}
    \resizebox{\textwidth}{!}{\begin{tabular}{ l | l }
        \toprule
        \multicolumn{1}{l|}{\textbf{Dataset}} &
        \multicolumn{1}{|c}{\textbf{Website (Link)}} \\
        \midrule
        MS MARCO & \url{https://microsoft.github.io/msmarco/} \\
        TREC-COVID & \url{https://ir.nist.gov/covidSubmit/index.html} \\
        NFCorpus & \url{https://www.cl.uni-heidelberg.de/statnlpgroup/nfcorpus/} \\
        BioASQ & \url{http://bioasq.org} \\
        NQ & \url{https://ai.google.com/research/NaturalQuestions} \\
        HotpotQA & \url{https://hotpotqa.github.io} \\
        FiQA-2018 & \url{https://sites.google.com/view/fiqa/} \\
        Signal-1M (RT) & \url{https://research.signal-ai.com/datasets/signal1m-tweetir.html} \\
        TREC-NEWS & \url{https://trec.nist.gov/data/news2019.html} \\
        Robust04 & \url{https://trec.nist.gov/data/t13_robust.html} \\
        ArguAna & \url{http://argumentation.bplaced.net/arguana/data} \\
        Touch\`e-2020 & \url{https://webis.de/events/touche-20/shared-task-1.html} \\
        CQADupStack & \url{http://nlp.cis.unimelb.edu.au/resources/cqadupstack/} \\
        Quora & \url{https://www.quora.com/q/quoradata/First-Quora-Dataset-Release-Question-Pairs} \\
        DBPedia-Entity & \url{https://github.com/iai-group/DBpedia-Entity/} \\
        SCIDOCS & \url{https://allenai.org/data/scidocs} \\
        FEVER & \url{http://fever.ai} \\
        Climate-FEVER & \url{http://climatefever.ai} \\
        SciFact & \url{https://github.com/allenai/scifact} \\
        \bottomrule
    \end{tabular}}
    \caption{Original dataset website (link) for all datasets present in \textbf{\beir}.}
    \label{tab:dataset_links}
\end{table*}

\begin{table*}[t]
    \small
    \ra{1.2}
    \resizebox{\textwidth}{!}{\begin{tabular}{ l | l }
        \toprule
        \multicolumn{1}{l|}{\textbf{Model}} &
        \multicolumn{1}{c}{\textbf{Public Model Checkpoints (Link)}} \\ \midrule
        BM25 (Anserini) & \url{https://github.com/castorini/anserini} \\
        DeepCT & \url{http://boston.lti.cs.cmu.edu/appendices/arXiv2019-DeepCT-Zhuyun-Dai/} \\
        SPARTA & \url{https://huggingface.co/BeIR/sparta-msmarco-distilbert-base-v1} \\
        DocT5query & \url{https://huggingface.co/BeIR/query-gen-msmarco-t5-base-v1} \\
        DPR (Query) &  \url{https://huggingface.co/sentence-transformers/facebook-dpr-question_encoder-multiset-base} \\
        DPR (Context) & \url{https://huggingface.co/sentence-transformers/facebook-dpr-ctx_encoder-multiset-base} \\
        ANCE & \url{https://huggingface.co/sentence-transformers/msmarco-roberta-base-ance-firstp} \\
        TAS-B & \url{https://huggingface.co/sentence-transformers/msmarco-distilbert-base-tas-b} \\
        ColBERT & \url{https://public.ukp.informatik.tu-darmstadt.de/thakur/BEIR/models/ColBERT/msmarco.psg.l2.zip} \\
        MiniLM-L6 (CE) & \url{https://huggingface.co/cross-encoder/ms-marco-MiniLM-L-6-v2} \\
        \bottomrule
    \end{tabular}}
    \caption{Publicly available model links used for evaluation in \textbf{\beir}.}
    \label{tab:model_links}
\end{table*}

\begin{table*}[t]
    \small
    \ra{1.2}
    \resizebox{\textwidth}{!}{\begin{tabular}{ l | l }
        \toprule
        \multicolumn{1}{l|}{\textbf{Corpus}} &
        \multicolumn{1}{|c}{\textbf{Website (Link)}} \\
        \midrule
        CORD-19 & \url{https://www.semanticscholar.org/cord19} \\
        NutritionFacts & \url{https://nutritionfacts.org} \\
        PubMed & \url{https://pubmed.ncbi.nlm.nih.gov} \\
        Signal-1M  & \url{https://research.signal-ai.com/datasets/signal1m.html} \\
        TREC Washington Post & \url{https://ir.nist.gov/wapo/} \\
        TREC disks 4 and 5 & \url{https://trec.nist.gov/data/cd45/} \\
        Args.me & \url{https://zenodo.org/record/4139439/} \\
        DBPedia (2015-10) & \url{http://downloads.dbpedia.org/wiki-archive/Downloads2015-10.html} \\
        TREC-COVID (Annotated) & \url{https://public.ukp.informatik.tu-darmstadt.de/thakur/BEIR/datasets/trec-covid-beir.zip} \\
        \bottomrule
    \end{tabular}}
    \caption{Corpus Name and Link used for datasets in \textbf{\beir}.}
    \label{tab:corpus_links}
\end{table*}

\begin{table*}[t!]
    \small
    \ra{1.1}
    \resizebox{\textwidth}{!}{\begin{tabular}{l | l | l }
        \toprule
        \multicolumn{1}{l|}{\textbf{Dataset}}    &
        \multicolumn{1}{c}{\textbf{Query}}   &
        \multicolumn{1}{|c}{\textbf{Relevant-Document}} \\
        \midrule
   MS MARCO & \multicolumn{1}{p{6cm}|}{what fruit is native to australia} & \multicolumn{1}{p{13cm}}{\textit{<Paragraph>} Passiflora herbertiana. A rare passion fruit native to Australia. Fruits are green-skinned, white fleshed, with an unknown edible rating. Some sources list the fruit as edible, sweet and tasty, while others list the fruits as being bitter and inedible. assiflora herbertiana. A rare passion fruit native to Australia...} \\ \midrule
   TREC-COVID    & \multicolumn{1}{p{6cm}|}{what is the origin of COVID-19} & \multicolumn{1}{p{13cm}}{\textit{<Title>} Origin of Novel Coronavirus (COVID-19): A Computational Biology Study using Artificial Intelligence \textit{<Paragraph>} Origin of the COVID-19 virus has been intensely debated in the community...} \\ \midrule
   BioASQ        & \multicolumn{1}{p{6cm}|}{What is the effect of HMGB2 loss on CTCF clustering} & \multicolumn{1}{p{13cm}}{\textit{<Title>} HMGB2 Loss upon Senescence Entry Disrupts Genomic Organization and Induces CTCF Clustering across Cell Types. \textit{<Paragraph>} Processes like cellular senescence are characterized by complex events giving rise to heterogeneous cell populations. However, the early molecular events driving this cascade remain elusive....}\\ \midrule
   NFCorpus      & \multicolumn{1}{p{6cm}|}{Titanium Dioxide \& Inflammatory Bowel Disease} & \multicolumn{1}{p{13cm}}{\textit{<Title>} Titanium Dioxide Nanoparticles in Food and Personal Care Products \textit{<Paragraph>} Titanium dioxide is a common additive in many food, personal care, and other consumer products used by people, which after use can enter the sewage system, and subsequently enter the environment as treated effluent discharged to surface waters or biosolids applied to agricultural land, or incinerated wastes...} \\ \midrule
   NQ            & \multicolumn{1}{p{6cm}|}{when did they stop cigarette advertising on television?} & \multicolumn{1}{p{13cm}}{\textit{<Title>} Tobacco advertising \textit{<Paragraph>} The first calls to restrict advertising came in 1962 from the Royal College of Physicians, who highlighted the health problems and recommended stricter laws...}\\ \midrule
   HotpotQA      & \multicolumn{1}{p{6cm}|}{Stockely Webster has paintings hanging in what home (that serves as the residence for the Mayor of New York)?}& \multicolumn{1}{p{13cm}}{\textit{<Title>} Stokely Webster \textit{<Paragraph>} Stokely Webster (1912 – 2001) was best known as an American impressionist painter who studied in Paris. His paintings can be found in the permanent collections of many museums, including the Metropolitan Museum of Art in New York, the National Museum...} \\ \midrule
    FiQA-2018    & \multicolumn{1}{p{6cm}|}{What is the PEG ratio? How is the PEG ratio calculated? How is the PEG ratio useful for stock investing?} & \multicolumn{1}{p{13cm}}{\textit{<Paragraph>} PEG is Price/Earnings to Growth.  It is calculated as Price/Earnings/Annual EPS Growth.  It represents how good a stock is to buy, factoring in growth of earnings, which P/E does not.  Obviously when PEG is lower, a stock is more undervalued, which means that it is a better buy, and more likely...} \\ \midrule
   Signal-1M (RT) & \multicolumn{1}{p{6cm}|}{Genvoya, a Gentler Anti-HIV Cocktail, Okayed by EU Regulators} & \multicolumn{1}{p{13cm}}{\textit{<Paragraph>} All people with \#HIV should get anti-retroviral drugs: @WHO, by @kkelland  via @Reuters\_Health \#AIDS \#TasP}\\ \midrule
   TREC-NEWS     & \multicolumn{1}{p{6cm}|}{Websites where children are prostituted are immune from prosecution. But why?} & \multicolumn{1}{p{13cm}}{\textit{<Title>} Senate launches bill to remove immunity for websites hosting illegal content, spurred by Backpage.com \textit{<Paragraph>}  The legislation, along with a similar bill in the House, sets the stage for a battle between Congress and some of the Internet’s most powerful players, including Google and various free-speech advocates, who believe that Congress shouldn’t regulate Web content or try to force websites to police themselves more rigorously...}\\ \midrule
   Robust04 & \multicolumn{1}{p{6cm}|}{What were the causes for the Islamic Revolution relative to relations with the U.S.?} & \multicolumn{1}{p{13cm}}{\textit{<Paragraph>}  BFN [Editorial: "Sow the Wind and Reap the Whirlwind"]  Yesterday marked the 14th anniversary of severing of diplomatic relations between the Islamic Republic and the United States of America. Several occasions arose in the last decade and a half for improving Irano-American relations...} \\ \midrule
   Touch\'e-2020 & \multicolumn{1}{p{6cm}|}{Should the government allow illegal immigrants to become citizens?} & \multicolumn{1}{p{13cm}}{\textit{<Title>} America should support blanket amnesty for illegal immigrants. \textit{<Paragraph>} Undocumented workers do not receive full Social Security benefits because they are not United States citizens " nor should they be until they seek citizenship legally. Illegal immigrants are legally obligated to pay taxes...} \\ \midrule 
   CQADupStack   & \multicolumn{1}{p{6cm}|}{Command to display first few and last few lines of a file} & \multicolumn{1}{p{13cm}}{\textit{<Title>} Combing head and tail in a single call via pipe \textit{<Paragraph>} On a regular basis, I am piping the output of some program to either `head` or `tail`. Now, suppose that I want to see the first AND last 10 lines of piped output, such that I could do something like ./lotsofoutput | headtail...} \\ \midrule
   Quora         & \multicolumn{1}{p{6cm}|}{How long does it take to methamphetamine out of your blood?} & \multicolumn{1}{p{13cm}}{\textit{<Paragraph>} How long does it take the body to get rid of methamphetamine?} \\ \midrule
   DBPedia       & \multicolumn{1}{p{6cm}|}{Paul Auster novels} & \multicolumn{1}{p{13cm}}{\textit{<Title>} The New York Trilogy \textit{<Paragraph>} The New York Trilogy is a series of novels by Paul Auster.  Originally published sequentially as City of Glass (1985), Ghosts (1986) and The Locked Room (1986), it has since been collected into a single volume.} \\ \midrule
   SCIDOCS       & \multicolumn{1}{p{6cm}|}{CFD Analysis of Convective Heat Transfer Coefficient on External Surfaces of Buildings} & \multicolumn{1}{p{13cm}}{ \textit{<Title>} Application of CFD in building performance simulation for the outdoor environment: an overview \textit{<Paragraph>} This paper provides an overview of the application of CFD in building performance simulation for the outdoor environment, focused on four topics...} \\ \midrule
   FEVER         & \multicolumn{1}{p{6cm}|}{DodgeBall: A True Underdog Story is an American movie from 2004} & \multicolumn{1}{p{13cm}}{\textit{<Title>} DodgeBall: A True Underdog Story \textit{<Paragraph>} DodgeBall: A True Underdog Story is a 2004 American sports comedy film written and directed by Rawson Marshall Thurber and starring Vince Vaughn and Ben Stiller. The film follows friends who enter a dodgeball tournament... } \\ \midrule
   Climate-FEVER & \multicolumn{1}{p{6cm}|}{Sea level rise is now increasing faster than predicted due to unexpectedly rapid ice melting.} & \multicolumn{1}{p{13cm}}{\textit{<Title>} Sea level rise \textit{<Paragraph>} A sea level rise is an increase in the volume of water in the world 's oceans, resulting in an increase in global mean sea level. The rise is usually attributed to global climate change by thermal expansion of the water in the oceans and by melting of Ice sheets and glaciers...}\\ \bottomrule
    \end{tabular}}
    \caption{Examples of queries and relevant documents for all datasets included in \textbf{\beir}. (\textit{<Title>}) and (\textit{<Paragraph>}) are used to distinguish the title separately from the paragraph within a document in the table above. These tokens were not passed to the respective models.}
    \label{tab:examples}
\end{table*}

\begin{table*}[t!]
    \ra{1.2}
    \resizebox{\textwidth}{!}{\begin{tabular}{l | c | c c c | c c c c | c | c}
        \toprule
        \multicolumn{1}{l}{\textbf{Model ($\rightarrow$)}} &
        \multicolumn{1}{c}{Lexical}   &
        \multicolumn{3}{c}{Sparse}   &
        \multicolumn{4}{c}{Dense} &
        \multicolumn{1}{c}{Late-Interaction} &
        \multicolumn{1}{c}{Re-ranking} \\ 
        \cmidrule(lr){1-1}
        \cmidrule(lr){2-2}
        \cmidrule(lr){3-5}
        \cmidrule(lr){6-9}
        \cmidrule(lr){10-10}
        \cmidrule(lr){11-11}
        \multicolumn{1}{l}{\textbf{Dataset ($\downarrow$)}} &
        \multicolumn{1}{c}{\textbf{BM25}} &
        \multicolumn{1}{c}{\textbf{DeepCT}} &
        \multicolumn{1}{c}{\textbf{SPARTA}} &
        \multicolumn{1}{c}{\textbf{docT5query}} &
        \multicolumn{1}{c}{\textbf{DPR}} &
        \multicolumn{1}{c}{\textbf{ANCE}} &
        \multicolumn{1}{c}{\textbf{TAS-B}} &
        \multicolumn{1}{c}{\textbf{GenQ}} &
        \multicolumn{1}{c}{\textbf{ColBERT}} &
        \multicolumn{1}{c}{\textbf{BM25+CE}} \\
        \midrule
   MS MARCO      & 0.658 & 0.752$^\ddagger$ & 0.793$^\ddagger$ & 0.819$^\ddagger$ & 0.552 & 0.852$^\ddagger$ & \textbf{0.884$^\ddagger$} & \textbf{0.884$^\ddagger$} & \underline{0.865$^\ddagger$} & 0.658$^\ddagger$ \\  \midrule \midrule
   TREC-COVID    & \underline{0.498$^\star$} & 0.347$^\star$ & 0.409$^\star$ & \textbf{0.541$^\star$} & 0.212$^\star$ & 0.457$^\star$ & 0.387$^\star$ & 0.456$^\star$ & 0.464$^\star$ & \underline{0.498$^\star$} \\
   BioASQ        & \textbf{0.714} & \underline{0.699} & 0.351 & 0.646 & 0.256 & 0.463 & 0.579 & 0.627 & 0.645 & \textbf{0.714} \\
   NFCorpus      & 0.250 & 0.235 & 0.243 & 0.253 & 0.208 & 0.232 & \textbf{0.280} & \textbf{0.280} & \underline{0.254} & 0.250 \\ \midrule
   NQ            & 0.760 & 0.636 & 0.787 & 0.832 & 0.880$^\ddagger$ & 0.836 & \underline{0.903} & 0.862 & \textbf{0.912} & 0.760 \\ 
   HotpotQA      & \underline{0.740} & 0.731 & 0.651 & 0.709 & 0.591 & 0.578 & 0.728 & 0.673 & \textbf{0.748} & \underline{0.740} \\ 
   FiQA-2018     & 0.539 & 0.489 & 0.446 & 0.598 & 0.342 & 0.581 & 0.593 & \textbf{0.618} & \underline{0.603} & 0.539 \\ \midrule
   Signal-1M (RT)& \textbf{0.370} & 0.299 & 0.270 & \underline{0.351} & 0.162 & 0.239 & 0.304 & 0.281 & 0.283 & \textbf{0.370} \\ \midrule
   TREC-NEWS     & \underline{0.422} & 0.316 & 0.262 & \textbf{0.439} & 0.215 & 0.398 & 0.418 & 0.412 & 0.367 & \underline{0.422} \\
   Robust04      & \textbf{0.375} & 0.271 & 0.215 & \underline{0.357} & 0.211 & 0.274 & 0.331 & 0.298 & 0.310 & \textbf{0.375} \\ \midrule
   ArguAna       & 0.942 & 0.932 & 0.893 & \underline{0.972} & 0.751 & 0.937 & 0.942 & \textbf{0.978} & 0.914 & 0.942 \\  
   Touch\'e-2020 & \underline{0.538} & 0.406 & 0.381 & \textbf{0.557} & 0.301 & 0.458 & 0.431 & 0.451 & 0.439 & \underline{0.538} \\ \midrule 
   CQADupStack   & 0.606 & 0.545 & 0.521 & \underline{0.638} & 0.403 & 0.579 & 0.622 & \textbf{0.654} & 0.624 & 0.606 \\
   Quora         & 0.973 & 0.954 & 0.896 & 0.982 & 0.470 & 0.987 & 0.986 & \underline{0.988} & \textbf{0.989} & 0.973 \\ \midrule
   DBPedia       & 0.398 & 0.372 & 0.411 & 0.365 & 0.349 & 0.319 & \textbf{0.499} & 0.431 & \underline{0.461} & 0.398 \\ \midrule
   SCIDOCS       & \underline{0.356} & 0.314 & 0.297 & \textbf{0.360} & 0.219 & 0.269 & 0.335 & 0.332 & 0.344 & \underline{0.356} \\ \midrule
   FEVER         & 0.931 & 0.735 & 0.843 & 0.916 & 0.840 & 0.900 & \textbf{0.937} & 0.928 & \underline{0.934} & 0.931 \\ 
   Climate-FEVER & 0.436 & 0.232 & 0.227 & 0.427 & 0.390 & 0.445 & \textbf{0.534} & \underline{0.450} & 0.444 & 0.436 \\  
   SciFact       & \underline{0.908} & 0.893 & 0.863 & \textbf{0.914} & 0.727 & 0.816 & 0.891 & 0.893 & 0.878 & \underline{0.908} \\
    \bottomrule
    \end{tabular}}
    \caption{In-domain and zero-shot retrieval performance on \beir datasets. Scores denote \textbf{Recall@100}. The best retrieval performance on a given dataset is marked in \textbf{bold}, and the second best performance is \underline{underlined}. $\ddagger$ indicates in-domain retrieval performance. $\star$ shows the capped Recall@100 score (Appendix \ref{sec:capped_recall_score}). \vspace{-5mm}}
    \label{tab:results-recall}
\end{table*}

\begin{table}[htb!]
    \small
    \includegraphics[trim=5 5 5 20,clip,width=\textwidth]{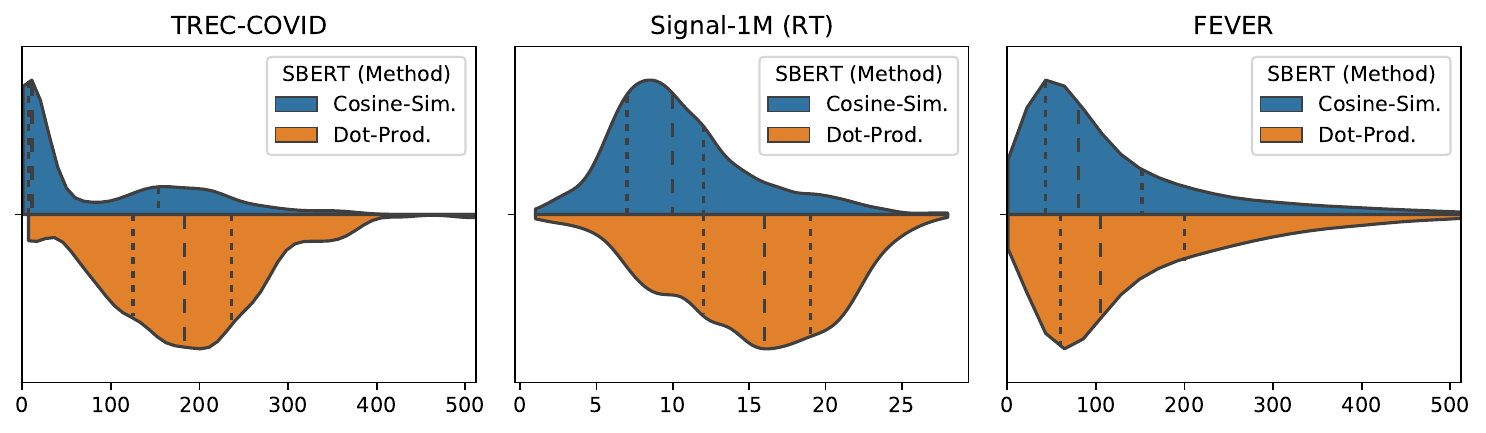}
    \resizebox{\textwidth}{!}{\begin{tabular}{c c c c c c}

        \multicolumn{2}{c}{\textbf{TREC-COVID}} &
        \multicolumn{2}{c}{\textbf{Signal-1M (RT)}} &
        \multicolumn{2}{c}{\textbf{FEVER}} \\
        \cmidrule(lr){1-2}
        \cmidrule(lr){3-4}
        \cmidrule(lr){5-6}
        \textbf{Cosine-Sim.} & \textbf{Dot-Prod.} & \textbf{Cosine-Sim.} & \textbf{Dot-Prod.} & \textbf{Cosine-Sim.} & \textbf{Dot-Prod.} \\ \midrule
        
        0.482 & \textbf{0.635} & \textbf{0.261} & 0.243 & 0.670 & \textbf{0.685} \\ \midrule
    \end{tabular}}
    \caption{Violin plots \cite{citeulike:4075875} of document lengths for the top-10 retrieved hits and nDCG@10 scores using a distilbert-base-uncased model trained with either cosine similarity (blue, top) or dot product (orange, bottom) as described in Appendix \ref{sec:dense_retrieval_length_preference}.}
    \label{tab:trec-covid}
    \vspace*{-\baselineskip}
\end{table}

\end{document}